\documentclass{raa07}
\usepackage{graphicx,times}
\usepackage{natbib}
\usepackage{url}
\usepackage{upgreek}
\usepackage{amssymb,amsmath}
\bibpunct{(}{)}{;}{a}{}{,}

\newcommand{\kms}{\,$\rm{km~s^{-1}}$}
\newcommand{\micron}{\,$\rm{\upmu m}$}

\newcommand{\degrc}{\ensuremath{\degr {\rm C}}}

\newcommand{\cntwo}{$C^2_n$}
\newcommand{\sunangle}{\ensuremath{\theta_{\tiny \sun}}}

\usepackage[a4paper=true,pagebackref=true]{hyperref}
\hypersetup{colorlinks = true, linkcolor = blue, anchorcolor = blue, citecolor = blue, filecolor = blue,  urlcolor = blue}

\begin{document}

   \title{Astronomy from Dome A in Antarctica}


\volnopage{{\bf 2020} Vol.~{\bf 20} No.~{\bf 10},~168(32pp)~
   {\small  doi: 10.1088/1674-4527/20/10/168}}
   \setcounter{page}{1}

   \author{Zhaohui Shang \thanks{\it{ORCID: 0000-0002-6796-124X}}}

   \institute{
National Astronomical Observatories, Chinese Academy of Sciences,
Beijing 100101, China; {\it zshang@gmail.com}\\
\vs \no
   {\small Received 2020 July 26; accepted 2020 September 17}
}

\abstract{  Dome~A in Antarctica has been demonstrated to be the best
site on earth for optical, infrared, and terahertz astronomical
observations by more and more evidence, such as excellent
free-atmosphere seeing, extremely low perceptible water vapor, low
sky background, and continuous dark time, etc. In this paper, we
present a complete picture of the development of astronomy at Dome~A
from the very beginning, review recent progress in time-domain
astronomy, demonstrate exciting results of the site testing, and
address the challenges in instrumentation.  Currently proposed
projects are briefly discussed. \keywords{ instrumentation:
miscellaneous
--- methods: observational
--- techniques: miscellaneous
--- telescopes
--- atmospheric effects
--- site testing
--- stars: variables: general
}
}

   \authorrunning{{\it Z. Shang}: Dome A Astronomy }            
   \titlerunning{{\it Z. Shang}: Dome A Astronomy }  
   \maketitle


%
\section{Introduction}   
\label{sect:intro}

Antarctic plateau has long been considered as an exceptional place for
ground-based astronomical observations because of its favorable
geographic and atmospheric properties.  It is often compared to space
and sometimes could be even better in terms of building large
observing facilities.

A huge polar ice cap covers most of the continent. Besides South Pole,
there are several prominent high points on the ice cap, among which
Dome~A is the highest at an altitude of 4093\,meters and located
within 80\degr\,S latitude circle.  The winter temperature drops to
between $-$50\degrc\ to $-$80\degrc\ while in summer it is seldom above
$-$30\degrc\ \citep{Hu14,Hu19}.

\begin{figure*}
   \centering
   \includegraphics[width=12.0cm, angle=0]{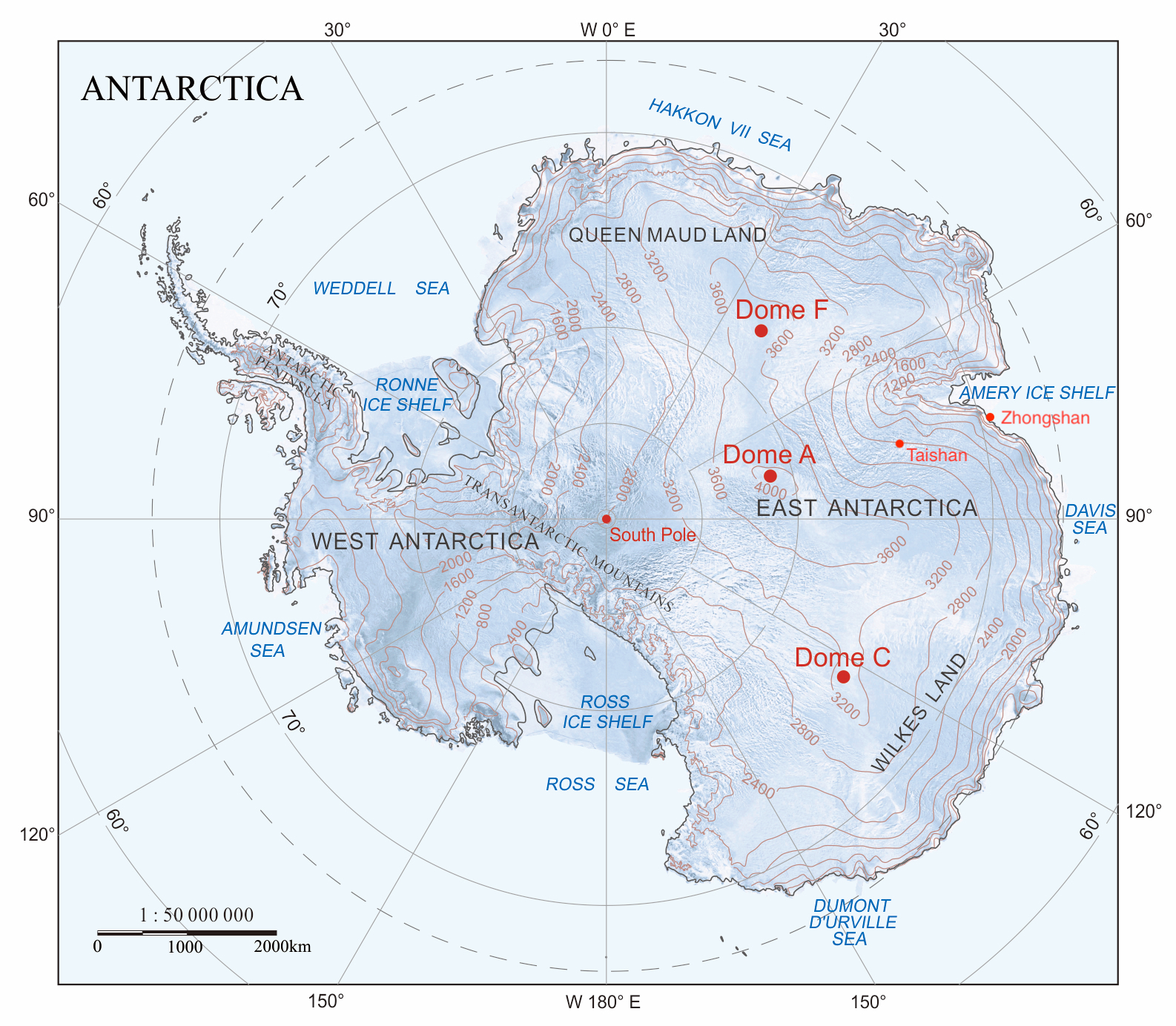}
   \caption{\baselineskip 3.8mm Map of Antarctica with elevation contour lines.
Marked on the map are Dome~A, C, F, and the South Pole as well as
Chinese Zhongshan and Taishan Stations. Courtesy of Xiaoping Pang and Shiyun Wang.}
   \label{fig:map}
\end{figure*}

Table~\ref{tab:domes} lists the
locations and altitudes of Dome~A, C, F, and South Pole which are also
marked on a map of Antarctica (Fig.~\ref{fig:map}).

\begin{table*}
\centering
\begin{minipage}[]{105mm}
\caption[]{Locations and Elevations of Some Sites on the Antarctic
Plateau\label{tab:domes}}
\end{minipage}

\setlength{\tabcolsep}{2pt}
\fns
 \begin{tabular}{lccrccl}
  \hline\noalign{\smallskip}
Site& ~ ~Latitude~ ~ && \multicolumn{1}{c}{Longitude} && Elevation &
\multicolumn{1}{c}{Station}\\
  \hline\noalign{\smallskip}
Dome A  & 80\degr22\arcmin\,S &&  77\degr21\arcmin\,E && 4093\,m
& Chinese Kunlun Station\\
Dome C  & 75\degr06\arcmin\,S && 123\degr20\arcmin\,E && 3233\,m
& French-Italian Concordia Station\\
Dome F  & 77\degr19\arcmin\,S &&  39\degr42\arcmin\,E && 3810\,m
& Japanese Dome Fuji Station\\
South Pole& --                &&  90\degr00\arcmin\,S && 2835\,m
& US Amundsen-Scott Station\\
  \noalign{\smallskip}\hline
\end{tabular}
\end{table*}

There are many advantages of doing astronomy in Antarctica,
especially at those high domes \citep{Storey05,Storey07,Ashley13}.
The air is super clean with minimal scattering light.  More
importantly,  temperature inversion near the ground often occurs,
resulting in a stable atmosphere when cooler air stays under warmer
air.  {W}hen coupled with low wind shear, {this
}can lead to very good astronomical seeing above a shallow surface
layer (or ground layer) of turbulence.

Moreover, the low temperature guarantees low thermal background
greatly benefiting infrared (IR) observations.
For sub-millimeter (sub-mm) wavelengths or terahertz (THz) frequencies
(1\,mm $\equiv$ 0.3\,THz), extremely cold weather results in extremely
dry air and thus low perceptible water vapor (PWV) which allows sub-mm
radiation to have a higher transmission rate through atmosphere and
reach ground more easily.
Another obvious advantage is the continuous dark time during polar
nights for time-domain astronomy which is becoming more and more
important.
The merits of potential observatory sites, including the ones in
Table~\ref{tab:domes}, were compared using limited measurements
available and satellite data \citep{Saunders09}.
There has been huge progress since then, especially for Dome~A.


South Pole station has hosted many astrophysics experiments mainly
in studies of cosmic microwave background (CMB) and neutrino{s}
since 1990s. The neutrino experiment IceCube is currently the
largest astronomical facility in Antarctica. There is also the
10-meter South Pole Telescope (SPT) for studies in microwave,
millimeter, and sub-mm wave bands, while BICEP and Keck Array are
specifically for CMB studies.
Early experiments in the infrared reported sufficiently low IR sky
background -- from 2.2\micron\ longwards the IR sky is between 20 to
100 times darker than that at a temperate latitude site
\citep{Ashley96,Phillips99}.
However, measured optical seeing is poor at South Pole with a mean
value of 1.8\arcsec\ \citep{Marks99,Travouillon03}.  Nevertheless, a
8\,cm small optical telescope SPOT was operated at South Pole station
from 1983 to 1988 \citep{Chen86,Taylor88},
but a proposed larger version (0.4\,m) does not seem to have ever been
built \citep{Chen92}.


At Dome C, a median optical seeing of 0.27\arcsec~was first reported
in 2004 \citep{Lawrence04} and later studies estimated it to be
0.23\arcsec--0.36\arcsec~above a boundary layer with a typical height of
30\,m \citep{Agabi06,Trinquet08,Aristidi09}.  Very low PWV level of
0.1--0.3\,mm were also found in winter \citep{Tremblin11}.
Along with the extensive site assessment, scientific studies have also
been carried out, mostly in time-domain astronomy.  Designed for
exoplanet search \citep{Daban10},   ASTEP-400 is a 40\,cm optical
telescope with a field-of-view (FOV) of 1~deg$^2$ and has found 43 transit exoplanet
candidates \citep{Mekarnia16} as well as variable stars
\citep{Mekarnia17}.
Its pathfinder, ASTEP-South \citep{Crouzet10}, has an aperture of only
10\,cm and a large FOV of $3.88\degr\times 3.88\degr$.  ASTEP-South
operated between 2008--2011 and ASTEP-400 worked from 2010 to 2013,
and both contributed to optical site testing as well as studies
of variable stars \citep{Crouzet16,Crouzet18}.
IRAIT, a 80\,cm infrared telescope,  was proposed \citep{Tosti06}
and started to work in 2013 with a mid-IR camera for 1.25--25\micron\
and a sub-mm (terahertz) camera for 200 and 350\micron\
\citep{Durand14}. The IRAIT also has a pathfinder, small-IRAIT,
working in the optical with an aperture of 25\,cm and
{an} FOV of $8\arcmin \times 5.3\arcmin$
\citep{Strassmeier08}.
Larger 2\,m class optical/infrared telescopes were also proposed for
Dome~C, such as PILOT \citep{Burton05} or PLT \citep{Epchtein11,Abe13}
which are yet to be realized.

Site testing at Dome~F has been conducted since 2000s.  Atmospheric
transparency at 220\,GHz was measured with a tipping radiometer and
a PWV of 0.6\,mm was reported for the 1-month period in 2006 summer
\citep{Ishii10}.  Optical seeing was also measured during the
austral summer of 2012/2013 for about 20 days, showing obvious
diurnal variation at a height of 11\,m.  Although the median seeing
value is 0.52\arcsec, very good seeing below 0.2\arcsec~was also
recorded \citep{Okita13}.
TwinCAM, consisting of two 10\,cm optical telescopes, was initially
installed at Dome~F for exoplanet search.  Larger experiments
involving a 40\,cm infrared and a 30\,cm THz telescopes were proposed
\citep{Ichikawa10}.  An even larger 2\,m class optical/infrared
telescope is under consideration \citep{Ichikawa16}.

Dome A, the highest location on the polar ice cap, was first visited
by humans in January 2005 via overland traverse from Zhongshan Station
by the 21st Chinese National Antarctic Research Expedition (CHINARE).
CHINARE is managed and carried out by Chinese Arctic and Antarctic
Administration (CAA) and Polar Research Institute of China (PRIC).
The distance between Dome~A and Zhongshan Station located at coast is
about 1200\,km (Fig.~\ref{fig:map}).
The Chinese astronomy community considered this as a viable
opportunity for ground-based astronomical research and reached an
agreement of eventually developing an observatory at Dome~A based on
its unique merits.  The Chinese Center for Antarctic Astronomy (CCAA)
was established in 2006 to coordinate relevant planning and
activities.  As the start, CCAA joined the second traverse to Dome~A
with the 24th CHINARE during the austral summer of 2007/2008.  Later,
the construction of Kunlun Station at Dome~A started in 2009 which
will be developed into a winterover station in the future.

During the past decade, there have been lots of developments at Dome~A
in site testing, astronomical research, as well as instrumentation for
harsh environment in Antarctica, all of which are reviewed in this
paper of the special issue (\citealt{Wang+Ip+2020}).  We start with the timeline of the development in astronomy at
Dome~A, together with introducing all the experiments and instruments
in Section~\ref{sec:timeline}.  We summarize the time-domain research
conducted at Dome~A in Section~\ref{sec:science} and review the results
of site testing in Section~\ref{sec:sitetesting}.  Instrumentation and
future plans are discussed relatively briefly at the end.
%
Earlier reviews before 2010 on astronomy and astrophysics from
Antarctica in general can be found in \citet{Storey05} and
\citet{Burton10}.

\section{Timeline and Instruments }
\label{sec:timeline}

The CHINARE has visited Dome~A every 1--2 years since its 2nd traverse
with scientific research in the areas of astronomy, glaciology,
subglacial geology, climatology, atmospheric physics, and surveying
and mapping, etc.  Hundreds of tons of fuel, construction materials and
scientific equipment were transported by an over-ice tractor convoy
and a team of usually about 20 members each time.  Given the
complications, it usually took about 20 days from Zhongshan Station to
Dome~A and the expedition team was able to work for only about 3 weeks
on-site until late January or early February before the temperature
dropped too low.  The Chinese astronomy community has joined every
traverse to Dome~A since the 24th CHINARE and managed to establish and
run an unattended, fully robotic observatory near Kunlun Station with
international collaborations.  All the devices and instruments have to
work fully automatically, with remote monitoring and communication via
Iridium satellites, until next traverse for necessary servicing.  This
has been a great challenge for instrumentation and operation.

Table~\ref{tab:traverse} summarizes astronomy-related facilities,
instruments, and devices during each CHINARE.  We try to compile the
information as complete as possible in order to serve as background
information as well as for future reference.

\begin{table*}
\centering
\begin{minipage}[]{90mm}
\caption[]{Timeline of Traverses and Instruments at
Dome~A\label{tab:traverse}}
\end{minipage}

\setlength{\tabcolsep}{1.5pt}
\fns
 \begin{tabular}{lcccc}
  \hline\noalign{\smallskip}
CHINARE \& Year &  ~ ~Members$^a$ & New Facilities \& Instruments &
Instruments removed \\
  \hline\noalign{\smallskip}
24th (2007/2008) & 2 &  PLATO, CSTAR, Pre-HEAT, Gattini, Snodar, DASLE  & \\
25th (2008/2009) & 1 &  Nigel, Snodar2, SAVER &  \\
26th (2009/2010) & 2 &  FTS, HRCAM, SHABAR & Snodar\\
27th (2010/2011) & 2 &  KLAWS, DIMM, SEU Platform & SHABAR, DIMM  \\
28th (2011/2012) & 4 &  AST3-1, PLATO-A, DIMM & CSTAR, Gattini, HRCAM, SEU
Platform, DIMM \\
29th (2012/2013) & 3 &  HRCAM2 & Nigel \\
30th (2013/2014)$^b$ & - &  -  & - \\
31st (2014/2015) & 2 &  AST3-2, KLAWS-2G, NIRSPEC, CSTAR-II & FTS \\
32nd (2015/2016) & 2 &  Webcams, Wind turbine & HRCAM2, NIRSPEC \\
33rd (2016/2017) & 4 &  KLCAM & AST3-1, PLATO, Snodar2, DASLE, KLAWS \\
34th (2017/2018)$^b$ & - &  -  & - \\
35th (2018/2019) & 4 &  KL-DIMM, nKLAWS-2G, KLCAM2, KLCAM3 & CSTAR-II,
Microthermal\\
 &  &  NISBM, MARST, Microthermal & \\
36th (2019/2020)$^b$ & - &  -  & - \\
  \noalign{\smallskip}\hline
\end{tabular}

\parbox{15.5cm}
{$^a$ Number of members in the traverse team from astronomy
community; $^b$ No Dome~A traverse.}
\end{table*}

\subsection{The 24th CHINARE}

\begin{figure*}
   \centering
   \includegraphics[width=14.0cm, angle=0]{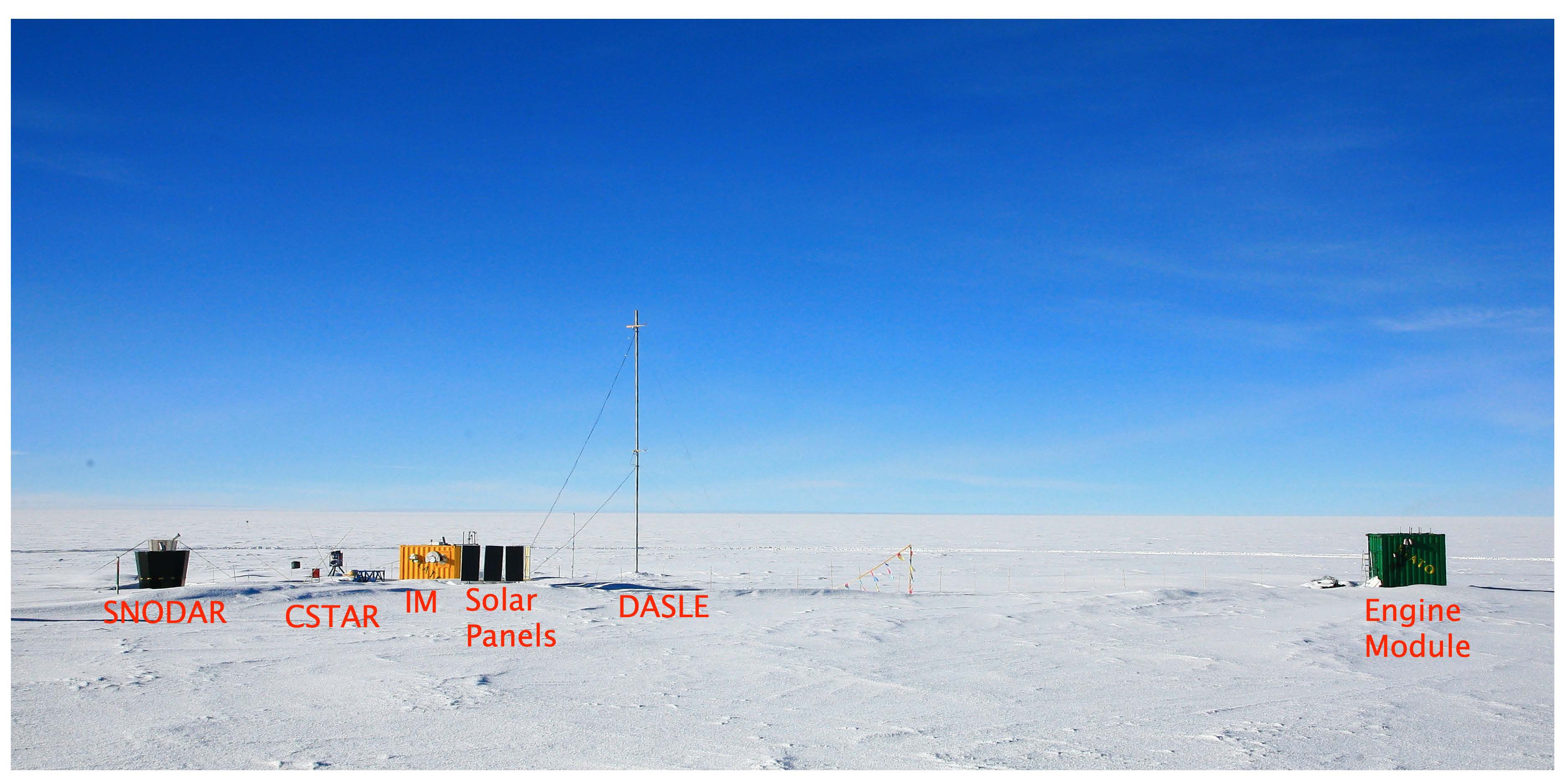}
   \caption{\baselineskip 3.8mm A photo showing the astronomy site and instruments in late
2009 during the 26th CHINARE.  The white instrument installed on the
side wall of the PLATO IM is Pre-HEAT.  Gattini was installed on the roof
of the IM.  The second Snodar ({\it silver}) is behind the first one ({\it black}).}
   \label{fig:site24}
\end{figure*}

This was the first time astronomers stepped into Dome~A.  With
collaborations of scientists from China, Australia, and the US, this
laid down the foundation of astronomical research at Dome~A.  The site
for astronomy was selected to be about 150\,m west of the planned Kunlun
Station to avoid possible contamination or adverse impacts in the
future.
An array of instruments for both site testing and astronomical observation
as well as supporting facilities were installed, as described below
(Fig.~\ref{fig:site24}).  A more detailed summary can be found in
\citet{Yang09}.

In the years followed, instruments had been serviced, upgraded, or
decommissioned, and new instruments had been installed until 2012 (the
28th CHINARE) when new site for astronomy was established.  The old
site was completely cleaned and all instruments removed in 2017 during
the 33rd CHINARE.

\subsubsection{PLATO}
\label{sec:plato}

Based on their expertise and experience from South Pole and Dome~C,
Australian scientists from the University of New South Wales (UNSW)
built for Dome~A the PLATeau Observatory (PLATO), which
is a self-contained automated facility that provides power and satellite
communications for itself and other instruments.
PLATO was designed to be extremely reliable and robust to withstand
the very low temperature, high altitude, high relative humidity, and
the constraint that servicing is only available at most once a year
\citep{Lawrence08,Luongvan08,Ashley10a}.

The main components of PLATO included an engine module (EM), an
instrument module (IM), and solar panels that supplemented the six
redundant diesel engines in the EM.  The IM housed a battery bank,
power supplies, and the control and communication computer systems,
such as the ``Supervisor'' computers.
It was also the place to host electronics and computers that operated
various instruments mounted either on the IM roof (or wall) ports or
externally placed on the snow surface.  The IM had an active thermal
management system and could keep the inside temperature well above
$-20$\degrc, higher than the working temperatures of customized
electronics and computers.  There were also webcams monitoring the
instruments and engines.

The EM was placed about 50\,m away from the IM to prevent the
instruments from being affected by exhaust stream or vibration.  PLATO
had a 4000-liter fuel tank in the EM, designed for continuous
operation through a winter and could in general provide power of 1\,kW
in average for everything.  This is a strong constraint and introduces
more challenges in developing instruments
(Sect.~\ref{sec:instrumentation}).

Currently, the communication at Dome~A can only be available via the
Iridium satellite network with a limited data allowance of hundreds of
MBytes per month depending on the data plan, beyond which the data
transferring becomes very expensive.
Necessary control and monitoring data are transferred to and from
Dome~A, but not the raw data from observations which have to be
retrieved until next traverse because of the data amount.  However,
the raw data can be processed on-site in real-time and the results,
with a much small data volume, can be downloaded so as not to delay
research.

\subsubsection{CSTAR}
\label{sec:cstar}

\begin{figure}
   \centering
   \includegraphics[width=7.0cm, angle=0]{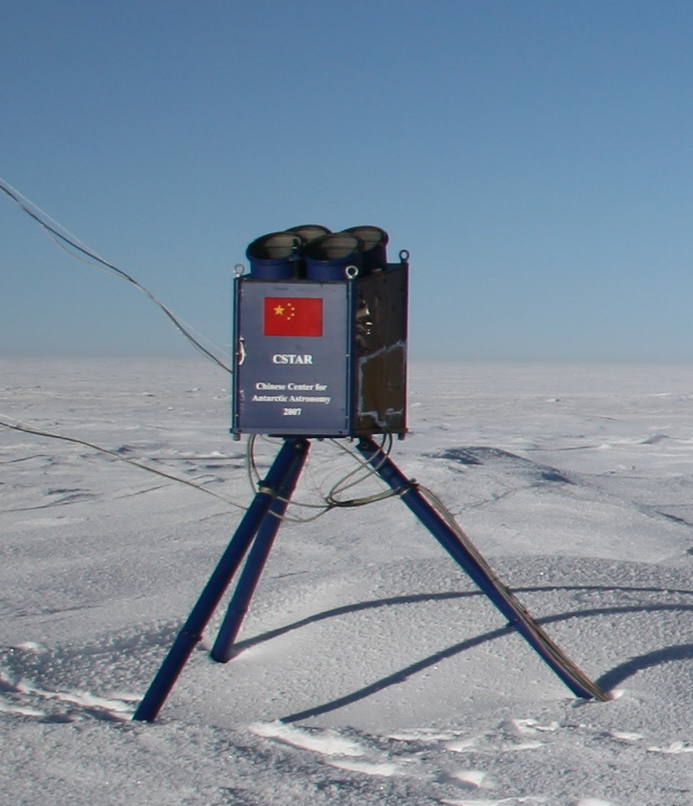}
   \begin{minipage}[]{85mm}
   \caption{\baselineskip 3.8mm CSTAR in late 2009 during the 26th CHINARE.}
   \label{fig:cstar}\end{minipage}
\end{figure}

Chinese Small Telescope ARray (CSTAR) was the first-generation
optical telescope at Dome~A (Fig.~\ref{fig:cstar}).  Its design
focused on simplicity and reliability, and eliminated any moving
parts, as it would be operated in an unknown world for the first
time \citep{Yuan08,Zhou10a,Zhou13}.  CSTAR consists of four
identical 14.5\,cm Schmidt-Cassegrain wide-field telescopes
(effective aperture of 10\,cm), for partial redundancy, with a large
focal ratio of f/1.2. Each telescope was equipped with an Andor
DV435 1k$\times$1k CCD camera with 13\micron\ pixels, resulting in a
plate scale of 15\arcsec~pixel$^{-1}$ and {an FOV} of about 20\,deg$^2$.  The CCD cameras were
operated in frame-transfer mode to eliminate the need of mechanical
shutters. Each telescope had a different filter: $g, r, i$, or
$clear$.  They were co-aligned and all pointed to the south
celestial pole (SCP). Each telescope had a separated sealing window
coated with Indium-Tin-Oxide (ITO) conductive film that can draw a
power of about 10\,W for defrosting when needed.

CSTAR was the first Chinese instrument and a key instrument for site
testing to monitor sky background, extinction (cloud), etc.  Its data
are also useful for time-domain astronomy.

CSTAR started to observe autonomously and continuously
{in} March 2008 and the exposure time was set to
10--30\,s.  Such a high cadence on a single field was unprecedented.
The data were retrieved during the next traverse and have been used
for studies of variable stars as well as evaluating the optical sky
background and extinction, etc. \citep{Zhou10b,Zou10}.  The data
have also showed that three of the four telescopes were out of focus
possibly because of the vibration during the journey to Dome~A, only
the $i$-band telescope remained well focused (but no data in 2009
for some reasons). Efforts were made to refocus the telescopes
on-site during the following visits to Dome~A, but without success,
because of the large focal ratio and the difficulties of working
during the daytime at Dome~A.  CSTAR was taken back in early 2012
during the 28th CHINARE for upgrading.

\subsubsection{Pre-HEAT}
\label{sec:preheat}

The first sub-mm instrument at Dome~A was Pre-HEAT
\citep{Kulesa08,Yang09}.  It had an aperture of 20\,cm and operated
at 600\,GHz (450\micron) with a Schottky diode heterodyne receiver
and a digital Fast Fourier Transform (FFT) spectrometer.  It was
installed on one side wall of PLATO (Fig.~\ref{fig:site24}),
measuring the 450\micron\ sky opacity and mapping the Galactic Plane
in the $^{13}$CO {$J$}=6-5 line at 661\,GHz with an
angular resolution of 10\arcmin\ \citep{Yang10}.  Pre-HEAT started
to work in January 2008 for one season.

Pre-HEAT was a technological prototype of the 60\,cm High Elevation
Antarctic Terahertz (HEAT) telescope which was later installed at
``Ridge~A'' in 2012, about 150\,km southwest of Dome~A
\citep{Walker04,Kulesa13}.

\subsubsection{Gattini}

Gattini was another optical instrument for monitoring the sky
brightness, cloud cover, aurora, and airglow
\citep{Moore08,Yangy17}. Earlier versions of Gattini had operated
successfully at Dome~C. Gattini at Dome~A had one narrow-field
Gattini Sky Brightness Camera (GSBC) and one wide-field Gattini
All-Sky Camera (GASC).  Each camera used an Apogee 2k$\times$2k
interline USB CCD array.  The GSBC used a Nikon lens with a 75\,mm
aperture and a focal length of 300\,mm (Nikon Telephoto AF-S Nikkor
300 mm f/4D ED-IF), resulting in a plate scale of
{5\arcsec~pixel$^{-1}$} and an FOV of
2.8\degr$\times$2.8\degr centered on the south celestial pole.  It
was equipped with Sloan filters $g', r'$ and $i'$.
The GASC used a Nikon wide-field lens with a focal length of
10.5\,mm, giving a pixel scale of
{145.4\arcsec~pixel$^{-1}$} and a large
FOV of 90\degr$\times$90\degr, looking toward zenith. The GASC was
equipped with Bessell filters $B, V, R$, and a longpass red filter
for the detection and monitoring of OH emission.

Gattini was installed on the roof of PLATO IM.  Due to some technical
problems, it did not obtain data for the first winter season.  After a
servicing mission in the next year, Gattini worked through the entire
winter of 2009 and was decommissioned with data retrieved during the
26th CHINARE.

\subsubsection{Snodar}
\label{sec:snodar1}

Snodar (Surface layer NOn-Doppler Acoustic Radar) was developed
specifically for Dome~C and Dome~A to measure the height and intensity
of the atmospheric boundary layer where most of the optical turbulence
exist on the Antarctic plateau.  Above the boundary layer,
free-atmosphere turbulence is extremely low and thus gives rise to
very good seeing.
Snodar is a monostatic high-frequency acoustic radar with a minimum
sampling height of 8\,m and a vertical resolution of 0.9\,m.  It
operated at frequencies between 3--15\,kHz and is sensitive to profile
optical turbulence to a height of 180\,m every 10\,s
\citep{Bonner08,Bonner10}.

Snodar worked for one week for the first season. A second Snodar
was deployed to Dome~A the next year and obtained excellent data
through the winter of 2009 \citep{Bonner10}.
%

\subsubsection{DASLE}

There was also the Dome A Surface Layer Experiment (DASLE) that was
designed to measure the temperature and 3D wind velocity, so as to
study the meteorological conditions, the intensity and vertical extent
of the boundary layer \citep{Yang09}.  DASLE had a mast of 17\,m tall with
fast sonic anemometers mounted at 4\,m, 8\,m, and 16\,m heights.

DASLE mast and sensors were successfully installed by the traverse
team and data were collected for one month.  The data showed that the
anemometer on the top level was damaged in transit and the other two
suffered ice formation due to problems with the power supply of the
deicing heater.  The problem was never fixed and DASLE was removed in
2017 with all other instruments on PLATO.

\subsection{The 25th CHINARE}
\label{sec:25th}

PLATO had worked autonomously for 204 days at Dome~A for the first
season, but stopped in mid-August 2008 before the traverse team
of the 25th CHINARE arrived in late 2008 for the servicing mission.

The team put a lot of efforts on maintaining PLATO and this has become
a routine for every following visit.  The maintenance included engine
replacement, oil change, and exhaust pipe replacement, etc. in the EM,
and necessary work on the computers, electronics, Iridium system,
battery bank, and solar panels, etc. on the IM side.  The batteries in the
IM needed to be replaced every 1-2 years.

CSTAR and Pre-HEAT were also maintained and Gattini was replaced.  A
second new Snodar was installed and the first one was repaired.  New
instruments and devices were also deployed (see below).

Construction of Kunlun Station began in 2009 during this traverse.

\subsubsection{Nigel}

Nigel was an optical/near-IR grating spectrometer operating in the
wavelengths of 300--850\,nm to monitor sky background, airglow and
aurorae \citep{Sims10,Sims12a}.  Nigel was fiber fed with no
additional optics, looking at three fixed positions: North at
40\degr~altitude, West at 71.5\degr~altitude, and the zenith.  A
stainless steel sphere with a diameter of 12\,cm held the fibers and
was mounted on the roof of PLATO IM.  Signals were fed through the
fibers to the spectrometer inside IM.  Each fiber has an FOV of
about 25\degr\ in diameter.  A thermoelectrically-cooled
256$\times$1024 pixel CCD camera is used to record the spectra with
a resolution of 3\,nm FWHM at 500\,nm (R$\sim$170).

\subsubsection{SAVER: shock and vibration recorder}

As the transit to Dome~A, especially during the overland traverse, is
very rough, some instruments and sensors had been damaged or become
defective.
A portable Lansmont SAVER 3X90 Field Data Logger was employed to
record the data in transit.  It had a magnetic mounting bracket and
could be easily installed on a shipping container with instruments.
The data collected over years have been very useful in planning future
transportation of instruments as well as instrument designs
\citep{Wen12}.

\subsection{The 26th CHINARE}
\label{sec:26th}

PLATO was still working when the traverse team arrived at Dome~A.
This demonstrated the reliability of the system and the feasibility of
running an unattended observatory remotely in Antarctica.  Routine
maintenance similar to the previous one was performed for PLATO
(Sect.~\ref{sec:25th}).

The team brought a collimator in another attempt to refocus
{three} of the {four} telescopes of CSTAR
during the daytime (Sect.~\ref{sec:cstar}), but data obtained in
winter showed that the problems were not entirely solved. A webcam
was installed to look at the windows of CSTAR, confirming snow
accumulation, instead of clouds, when the extinction of images
became large.

Gattini and the second Snodar were maintained while the first Snodar
was decommissioned.  Pre-HEAT stopped working sometime and was
decommissioned, but not uninstalled.

\begin{figure*}
   \centering
   \includegraphics[width=10.0cm, angle=0]{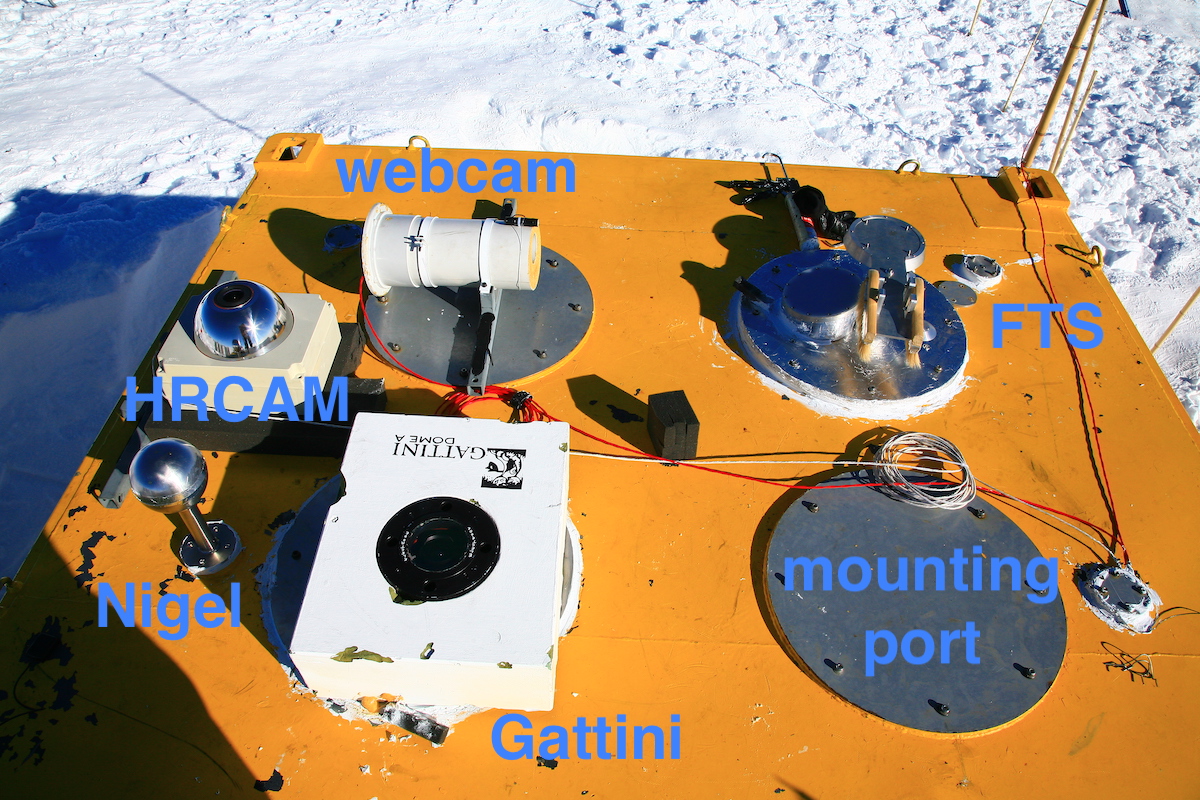}
  \begin{minipage}[]{170mm}
   \caption{\baselineskip 3.8mm Part of the roof of PLATO IM, showing some instruments
maintained or newly installed during the 26th CHINARE.}
   \label{fig:site26}\end{minipage}
\end{figure*}

To prepare for future Antarctic Survey Telescopes (AST3,
Sect.~\ref{sec:ast3}), the team selected a new site for astronomy,
further away from Kunlun Station and about 250\,m west of the current
site.  Since AST3 telescopes are much larger and heavier than CSTAR,
they could not be simply put on snow surface.
A 20\,m$\times$20\,m foundation was prepared for AST3 by digging to a
depth of 1.5\,m in average, backfilling to make the snow harder and
increase the capacity of the foundation (Yuansheng  Li, {\it private
communications}), and finally using a 23-ton heavy tractor as a roller
compactor.  The first AST3 telescope was installed at this site two
years later.
During this traverse, some new instruments were installed as described
below (Fig.~\ref{fig:site26}).

\begin{figure*}
   \centering
   \includegraphics[width=10.0cm, angle=0]{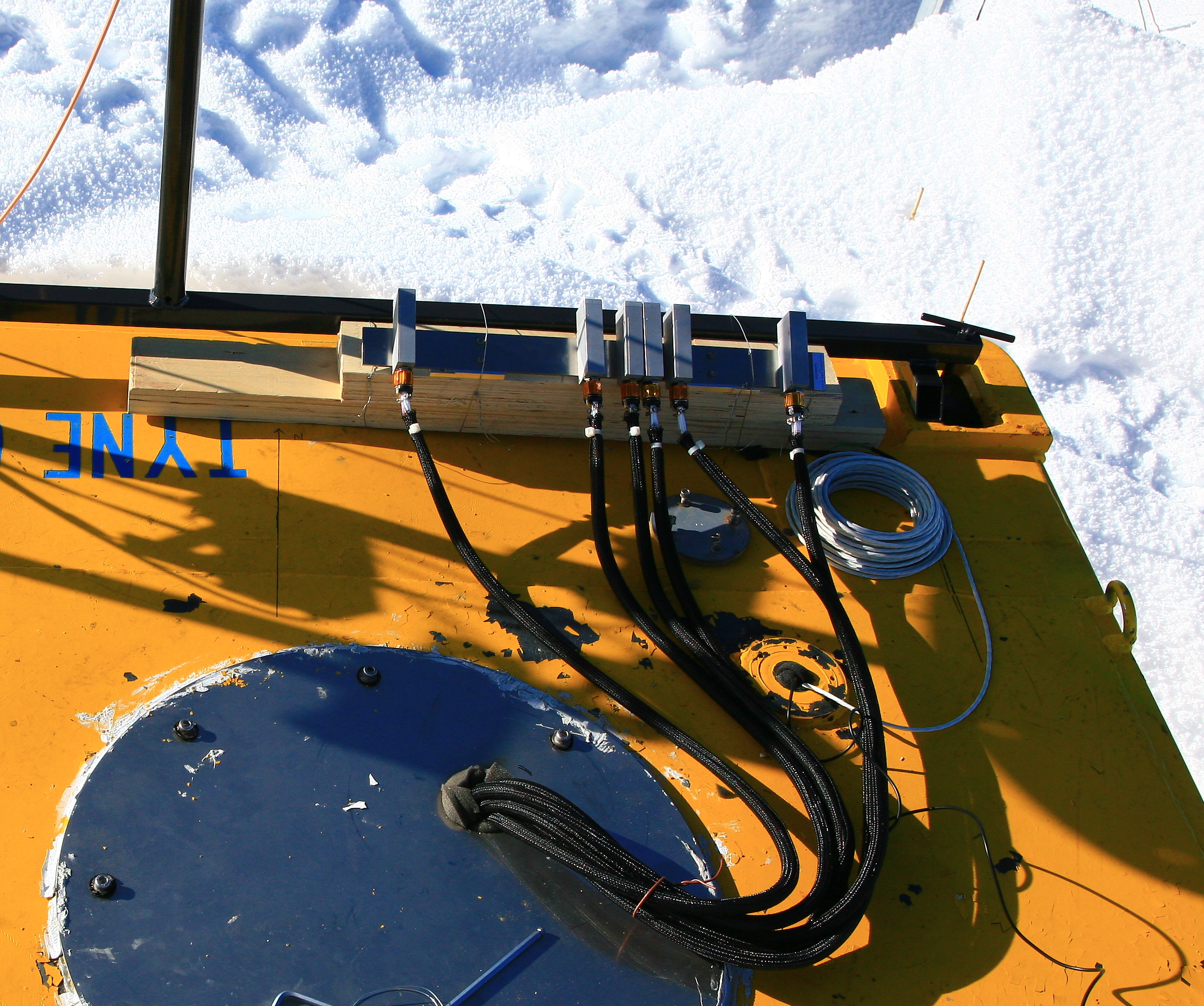}
   \begin{minipage}[]{110mm}
   \caption{\baselineskip 3.8mm Dome~A SHABAR was installed at an edge of the roof of PLATO
IM.}
   \label{fig:shabar}\end{minipage}
\end{figure*}

\subsubsection{FTS}
\label{sec:fts}

The Fourier Transform Spectrometer (FTS) was a polarizing Michelson
interferometer working in terahertz regime 750\,GHz -- 15\,THz with
an instrumental resolution of 13.8\,GHz \citep{Shi16}.  It measured
atmospheric radiation covering the entire water vapor pure
rotation band from 20\micron\ to 350\micron\ and reveal the
atmospheric transmission throughout the band.
The FTS was installed on one roof port of PLATO IM
(Sect.~\ref{sec:plato}), with the observing window and calibration
blackbody, etc. on top of the roof and the interferometer mounted on
the ceiling inside.  A webcam was installed on the roof to monitor the
operation of the external moving parts --- while the blackbody rotated
in and out above the window for calibration, a brush on the other side
could also clean the snow/frost off the window.

FTS operated from January 2009 for two seasons of a total 19 months
before it was decommissioned.

\subsubsection{HRCAM}
\label{sec:hrcam}

The High Resolution CAMera (HRCAM) was an all-sky camera, built with
a Canon EOS 50D digital SLR camera and a Sigma 4.5-mm f/2.8 fish-eye
lens \citep{Sims13}.  It took images every few minutes with a full
sky coverage, and the cadence was mostly limited by the lifetime of
the camera's shutter.  HRCAM was placed on the roof of PLATO IM
(Fig.~\ref{fig:site26}).  {A} special enclosure was
designed for the camera, lens, {an} ARM-based computer,
and electronics to cope with the low temperature down to
$-$80\degrc.
The data from HRCAM are useful for evaluating the fractions of time of
clouds and aurorae, however, the data did not cover a period long
enough for statistics during the two seasons of operation.

\subsubsection{SHABAR}
\label{sec:shabar}

SHABAR stands for SHAdow-BAnd Ranger, an instrument to measure
scintillation from an extended source such as the Sun
\citep{Beckers01,Tokovinin07}.  When it is used for the Moon, it is
also referred as Lunar Scintillometer \citep{Hickson04}.
It uses multiple photo-diodes arranged in a linear configuration.
Correlated intensity fluctuations at any two detectors reflects the
turbulence at a height corresponding to the baseline of the two
detectors.  With different baselines, SHABAR can profile the vertical
structure of the turbulence.

The Dome~A SHABAR had six photo-diodes attached to a single Invar
base plate with a maximum baseline of 40\,cm.  It was installed at
an edge of the IM roof facing north, with its electronics inside
PLATO ({F}ig.~\ref{fig:shabar}).  It shared a computer
with CSTAR and was tested to function properly. Some data were
collected in winter when the Moon was up, but it started to have
problems in July 2010.
It was found the next year that the windows in front of the
photo-diodes were broken, possibly due to the low temperature in
winter.  Other problems were not solved on-site and SHABAR was
therefore uninstalled.

\subsection{The 27th CHINARE}
\label{sec:27th}

PLATO worked through the winter again and the team serviced it and
replaced necessary parts like before for next year. CSTAR, Gattini,
Nigel, and Snodar were also serviced, and SHABAR was checked and
decommissioned because of problems that could not be fixed.  A full
year data was retrieved from FTS which continued to work well.

The team brought a DIMM in an attempt of daytime seeing measurement
\citep{Pei10,Pei12}, but was not successful due to technical issues.

The KunLun Automatic Weather Station (KLAWS) and a new facility, SEU
Platform, developed by CCAA and Southeast University (SEU), were
deployed during this traverse.

\subsubsection{KLAWS}
\label{sec:klaws}

KLAWS was a collaborative scientific project involving CCAA, UNSW, and
PRIC.  It was not a tradition weather station for meteorology
studies, it had a 15\,m-tall mast with a higher vertical resolution of
temperature and wind --- temperature sensors at nine heights from
$-$1\,m to 14.5\,m, and propeller anemometers at four heights
\citep{Hu14}.  This was designed to profile the low boundary layer at
Dome~A.

KLAWS began operation as soon as it was set up, performed well for a
full year, and collected very valuable data for us to understand
Dome~A, especially in winter.  The traverse team of the 28th CHINARE
found that its mast fell partially like an arch and they were not able
to fix it.  It is also believed that its power and data cables were
damaged accidentally and unknowingly before the team left.
A web page was developed to show the weather condition at Dome~A in
real-time and historical data from KLAWS are also
available\footnote{\it \url{http://aag.bao.ac.cn/weather/}}.

\subsubsection{SEU platform}

As more instruments were installed at Dome~A, demands for power and
logistic support became larger, especially with considerations of
future development of the observatory.
SEU Platform was designed as a PLATO-like self-contained facility of
similar size, providing 1\,kW power in average and communication via
Iridium network for unattended instruments at Dome~A.

Before being deployed to Dome~A, the platform was tested for a few
months at the remote Yangbajing, Tibet (4300\,m altitude) for
low-pressure condition, automated operation, and its overall
performance.
It was then successfully installed at Dome~A and operated for more
than 50 days before it stopped unfortunately due to a problem, likely
a software glitch as concluded later ({\it private communications}).
It was uninstalled from Dome~A the next year and brought back to
Zhongshan Station, and back to CCAA after a few years.
New versions of the platform have been tested at Taishan Station.

\subsection{The 28th CHINARE}
\label{sec:28th}

\begin{figure*}
   \centering
   \includegraphics[width=5.0cm, angle=0]{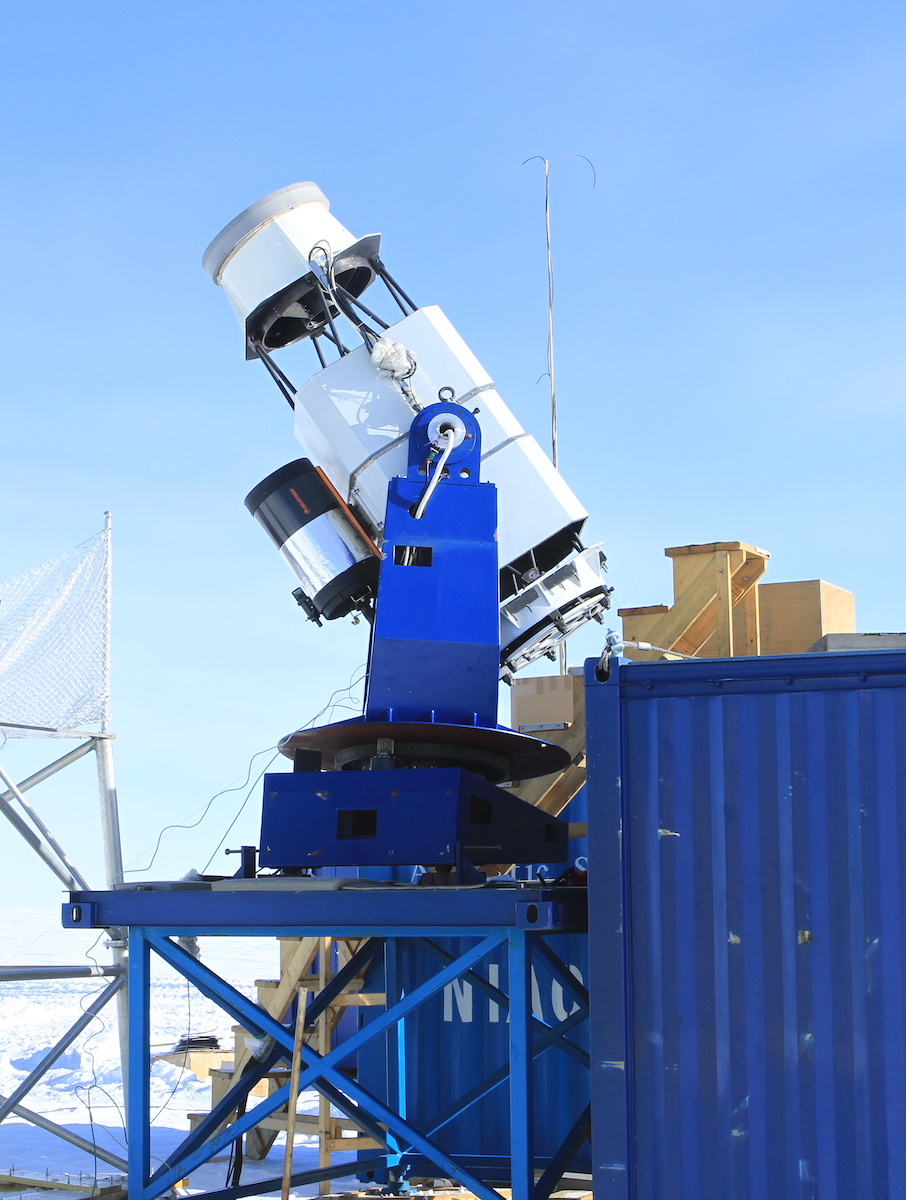}
   \includegraphics[width=5.0cm, angle=0]{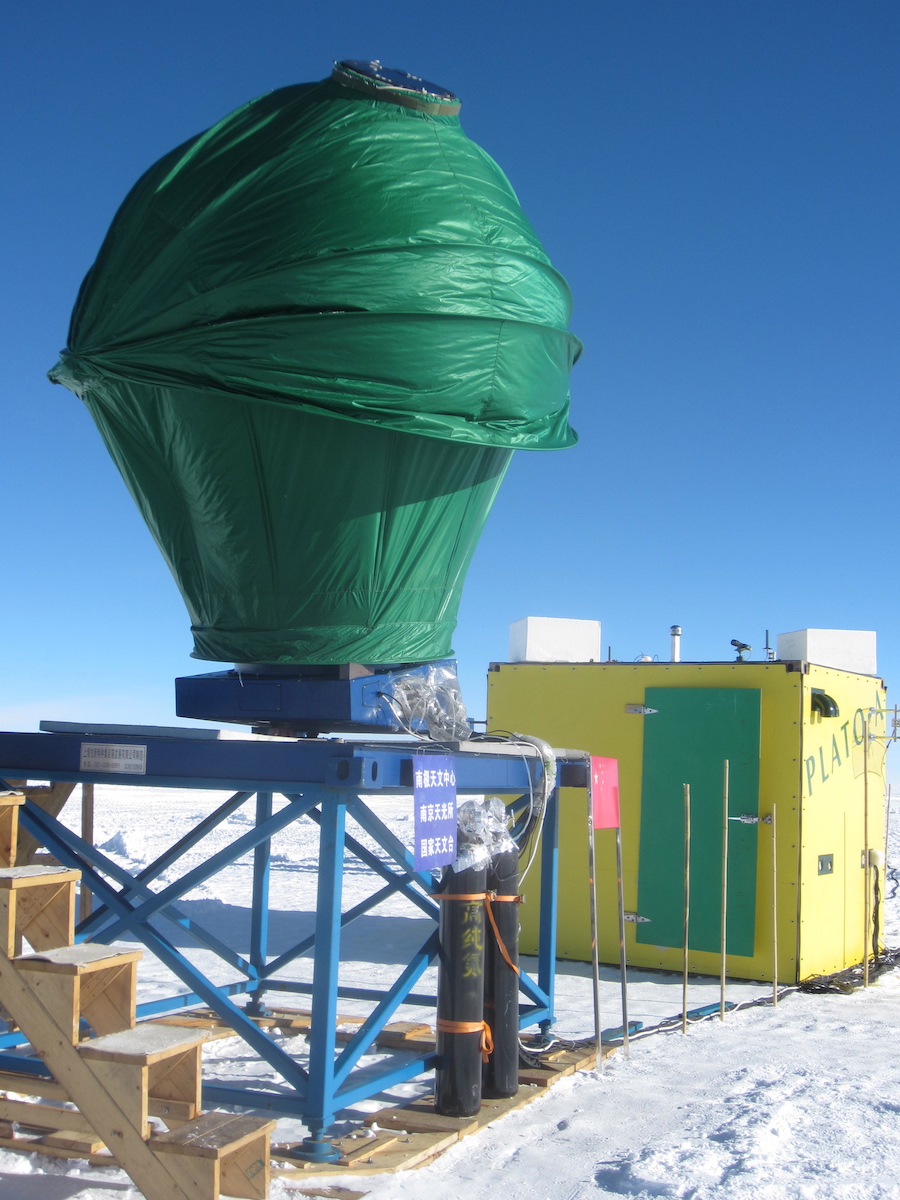}
   \caption{\baselineskip 3.8mm AST3-1 during its installation without and with the dome.
The small telescope tube attached to AST3-1 is the DIMM.
Courtesy of Fujia Du and Zhengyang Li.}
   \label{fig:ast3}
\end{figure*}

\begin{figure*}
   \centering
   \includegraphics[width=14.0cm, angle=0]{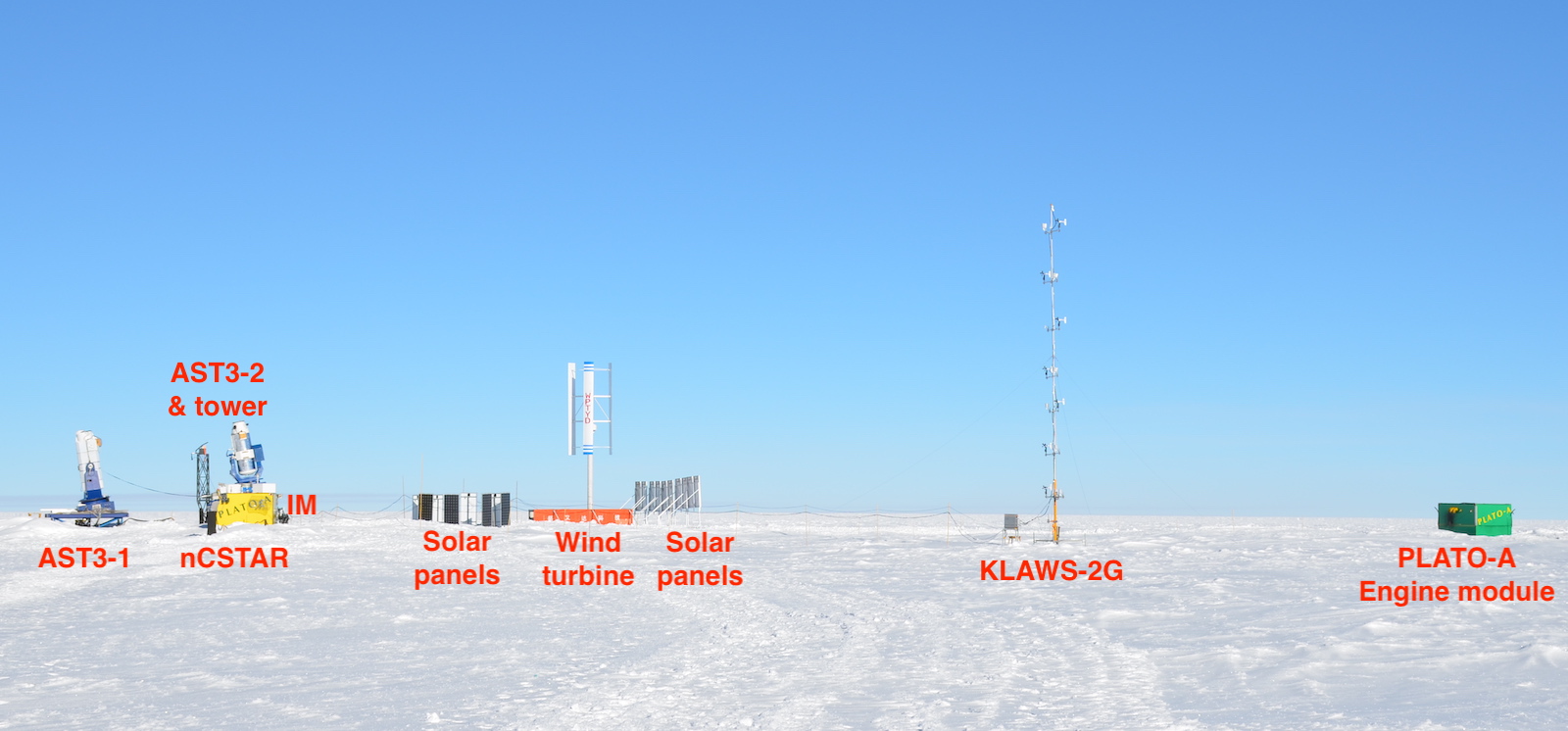}
   \caption{\baselineskip 3.8mm A photo showing the instruments at the new astronomy site
in early 2016 during the 32nd CHINARE. The CSTAR-II is small and in
front of the AST3-2 tower.
Courtesy of Ce Yu.
}
   \label{fig:site32}
\end{figure*}

This expedition and traverse marked another milestone of astronomy at
Dome~A.  The first Antarctic Survey Telescope AST3-1
(Fig.~\ref{fig:ast3})  was installed
and operated at Dome~A on the new site for astronomy, 240\,m west of
the old site (Sect.~\ref{sec:26th}).  A new supporting facility PLATO-A was
also deployed with AST3-1 (Fig.~\ref{fig:site32}).

The team brought a DIMM tube again for daytime seeing measurements
\citep{Pei12}.  It was installed on the larger telescope tube of
AST3-1, sharing the mount.  They managed to obtain daytime seeing
measurements for the first time for 3 days (Sect.~\ref{sec:seeing}).


At the old site, the mast of KLAWS fell partially
like an arch and could not be fixed.  Its power and data cables were
also damaged later (Sect.~\ref{sec:klaws}).
FTS and Snodar, as well as the old PLATO, were maintained, but did not
really work again for some reasons.  The old site for astronomy was
virtually decommissioned during this traverse.

The team uninstalled CSTAR, Gattini, and the SEU Platform and brought
them back.

\subsubsection{The AST3 project}
\label{sec:ast3}

AST3 is the second-generation optical telescope project of CCAA.  The
project was designed to run three identical 50\,cm telescopes
\citep{Cui08} with different filters simultaneously, so as to obtain
multi-color light curves for early SN discovery, exoplanet search, and
other time-domain studies (Fig.~\ref{fig:ast3}).

The AST3 telescopes, with full pointing, tracking, and focusing
capabilities, have a modified Schmidt design, achieving a wide FOV as
well as a good image quality with 80 per cent of a point source's
energy encircled within 1\arcsec.   Each telescope
has an entrance pupil diameter of 50\,cm and a focal ratio of f/3.73
\citep{Yuan10,Yuan12}.

Each telescope is equipped with an STA1600FT CCD camera built by
Semiconductor Technology Associates, Inc (STA).  Each camera has a
single-chip CCD of 10560$\times$10560 9\micron\ pixels, giving a plate
scale of 1\arcsec~pixel$^{-1}$.  To minimize mechanical failure in winter,
there is no shutter for the camera.  Instead, the camera is operated
in frame-transfer mode, using the top and bottom 10560$\times$2640
areas as buffers and the central 10560$\times$5280 pixels as the
exposing area.  This results in an effective FOV of
2.92\degr$\times$1.46\degr\ (4.26\,deg$^2$).  The CCD cameras went
through detailed lab tests in order to thoroughly understand their
performance as they cannot be easily accessed once they were deployed
to Dome~A with the telescopes \citep{Ma12}.  A rare problem was found
and studied that the photon transfer curve became non-linear at a
level around just 25000\,ADU, causing the brighter-fatter effect, and
could affect photometry of bright sources \citep{Ma14a}.
Other special treatments have also been studied for data reduction of
AST3 images \citep{Ma14b,Wei14} and will be discussed in
Section~\ref{sec:photo}.

A dedicated and customized Control, Operation and Data System (CODS)
was developed to run the unattended AST3 automatically through winter
at Dome~A \citep{Shang12,Shang16}.  CODS had three subsystems: a main
control system ({\it MAIN}) a data storage array ({\it ARRAY}) and a
pipeline system ({\it PIPE}).  Each subsystem has an identical backup
for redundancy (Sect.~\ref{sec:instrumentation}).  Over years, a suite
of software has also been developed and optimized for AST3, such as
survey planning \citep{Liuq18}, real-time photometry pipeline, etc., as
well as a sophisticated operation management system that can be used
for any robotic observatory \citep{Hu16,Hu18}.

As designed for AST3 to run at Dome~A, everything for the operation,
from observation planning to data reduction, had to be fully
automated.  To do so, CODS communicates, through {\it MAIN}, to the
telescope via a native software that actually drives the telescope
\citep{Lixy12,Lixy13}, takes exposures with CCD, distributes raw
images to {\it ARRAY} for permanently storage and to {\it PIPE} for
real-time photometry, and sends transit candidates and alerts back to
NAOC to display on a web page for further confirmation and decisions on
follow-up observations.  A more detailed description of the survey
automation and data reduction can be found in \citet{Ma20b}.

AST3-1 had an $i$-band filter and the second telescope AST3-2,
installed in 2015 during the 31st CHINARE (Sect.~\ref{sec:31st}), was
equipped with two filters ($i$ and $R$-bands).
For the future, the third telescope AST3-3 has a modification by
adding a fold mirror to move the focal plane outside the tube.  This
could largely solve the problem of CCD heat dissipation to improve tube
seeing (Sect.~\ref{sec:ast32}), simplify the camera mounting and
maintenance, and give the possibility of adding an IR camera
\citep{Lawrence16}.

\subsubsection{AST3-1}
\label{sec:ast31}

The functionality and operation of AST3-1 {were}
confirmed during a testing period of a couple of months at Xuyi
Station of Purple Mountain Observatory (PMO) right before it was
deployed to Dome~A. An altazimuth mount was used for AST3-1 during
the test because its equatorial mount was designed specifically for
the latitude of Dome~A. Very short time exposures showed good and
uniform image quality, demonstrating that the optical system was
assembled and aligned well \citep{Lizy12}.  Data reduction and
analysis of the first AST3 images was also done.  The camera of
AST3-1 had an engineering-grade CCD as no science-grade CCD was
available by STA then.  The camera was replaced with one with a
science-grade CCD three years later during the 31st CHINARE.

AST3-1 was installed at Dome~A in January 2012 on a 1.5\,m-tall
platform (Fig.~\ref{fig:ast3}) placed on the foundation prepared
during the 26th CHINARE.  The top of the telescope can reach up to
4\,m.
Having developed a new method using no star during the daytime
\citep{Lizy12,Lizy14}, the team did an excellent job aligning the
telescope and its optics.

A light, tent-like foldable dome covered the body of the telescope,
leaving the mirror outside, and moved with the telescope
\citep{Yuan10}.  The dome was designed to keep snow away from the
telescope, however, it seemed to be problematic for telescope
tracking and moving when there were gusts.  It was adjusted the next
year and was eventually removed during the 32nd CHINARE when the
problem was further confirmed.

A first version of CODS was also deployed to conduct the observing.
The commissioning started in early March 2012 when there was twilight
and confirmed the good image quality and alignment of the telescope.
The polar axis of the telescope was only off by 0.7\degr\ which was
small compared to its large FOV, and the pointing was improved to
better than tens of arcseconds after a pointing model was established
with observing data and the {\sc
tpoint}\footnote{\it \url{http://www.tpointsw.uk/}} software.

Later it seemed that the mechanical part of the telescope suffered
some problems --- the telescope could stop moving sometimes in the
declination (Dec) direction during tracking while the right ascension
(RA) axis worked well.  This was attributed to possible icing in the
gears and also some deformation of the Dec axis during transportation
\citep{Yuan14}.  Therefore, although CODS was designed to run an
automated survey, manual operation had to be done remotely from NAOC
and a team of more than 20 people from AST3 collaboration were
involved in the 24-hour non-stop operation.  Valuable data were
obtained for the first season for about 50 days
until early May \citep{Ma18a}  before the operation unfortunately
stopped due to a power supply problem to the telescope.

\subsubsection{PLATO-A}

PLATO-A is an optimized version of PLATO (Sect.~\ref{sec:plato})
with only {five} engines in the engine module
\citep{Ashley10b}. The functionality and size, etc. did not change,
but the new design of EM allows an easier access to the improved
engine systems for maintenance.  The size of the fuel tank increased
to 6000 liters. It was maintained by every traverse team and has
been supporting AST3 and other instruments to the date.

\subsection{The 29th CHINARE}
\label{sec:29th}

Most of the devices in the instrument module of PLATO-A were still
working when the team arrived.  PLATO-A was serviced just like for
PLATO (Sect.~{\ref{sec:25th}).

AST3-1 was checked and serviced, and its tent-like dome was
re-adjusted.  A new set of CODS system replaced the old one, and
AST3-1 data from last season were retrieved.
However, it was found later that the CCD controller did not work and
thus only engineering work was done with the telescope for the season
until more problems emerged in winter \citep{Yuan14}.

A new HRCAM was installed on the roof of PLATO-A IM.  Nigel from
old PLATO was decommissioned.

\subsection{The 31st CHINARE}
\label{sec:31st}

The expedition team returned to Dome~A after two years with a very
ambitious plan.  The second telescope AST3-2 was installed and
operated.  The second generation weather tower, KLAWS-2G, was
established as well as a near-IR spectrometer NIRSPEC.  CSTAR was
brought back as a new instrument CSTAR-II.

PLATO-A and AST3-1 were serviced and revived.  It was found that the
stuck problem with its Dec axis {was} caused by its
dome which was then removed, and one of the two motors which was
also removed. Its CCD camera controller was dead and it was upgraded
with a new controller and a new camera with a science-grade CCD.

FTS was uninstalled and returned to PMO, data were retrieved.

\subsubsection{AST3-2 and AST3-1}
\label{sec:ast32}

AST-2 is almost identical to AST3-1.  It was first tested at
Xinglong Station of NAOC on an altazimuth mount that did not allow
CODS to automate sky survey (planning, observing and data reduction,
etc.) like at Dome~A.  Later during {November} 2013 to
{April} 2014, AST3-2 was winterized at Mohe, the
coldest place in northeast China. With a customized mounting
structure, the telescope was able to run on an equatorial mount and
the automated operation at Dome~A was simulated.  It was
demonstrated for the first time that the system could carry out the
automated sky survey around the clock with real-time data reduction
and transient detection.

However, the weather situation at Mohe was still very different from
that at Dome~A.  The lowest temperature was only $-$40\degrc\ here
compared to $-$70\degrc\ to $-80$\degrc\ at Dome~A.  Frosting on the
front mirror was seen and different methods had been studied for
defrosting, but suspected icing on gears was not found.

AST3-2 was installed successfully near AST3-1 (Fig.~\ref{fig:site32})
and both telescopes worked well after the team left Dome~A.  Lots of
engineering tests on the telescopes were done remotely with when solar
power was enough and they got stuck occasionally.  The new CCD camera
for AST3-1 had problems in TEC cooling and configurations as it was
shipped in a rush.

As it got dark and the temperature dropped,  the two telescopes could
still work, but not smoothly. Engineering tests still dominated,
because more and more problems emerged with electronics, computers,
and defrosting, etc.
Also as the Sun went down, solar power decreased.  It was a mistake
that some instruments/devices were not designed with careful
consideration of minimizing power consumption.  PLATO-A was not
designed to provide more than 1\,kW in average.  In order to pursuing
overall success for the season, decisions had to be made in early
April to keep AST3-2, and shut down AST3-1 and CSTAR-II based on their
performance and possibility of getting scientific data.

Even for just AST3-2, along with other problems, frosting on the front
mirror became more and more severe.  To defrost, the ITO coating on
the mirror was found to be not enough and warm air from a blower
inside the tube was used from time to time.  This resulted in very
large tube seeing and bad image quality.  In May, the blower failed
and extinction on images became larger and larger as frost gradually
covered the whole mirror.
The season was ended without starting automatic operation and
everything was done manually as for AST3-1 three years before
(Sect.~\ref{sec:ast31}).  Exoplanet search was attempted at a time, but
there were no real scientific data obtained.
A lot of lessons have been learned from this season.

\subsubsection{KLAWS-2G}
\label{sec:klaws2g}

Weather information is critical for running an observatory, especially
for unattended instruments like AST3.  KLAWS-2G was built for this
purpose and was very useful in guiding the operation of AST3 this
season.
It was also used to monitor the boundary layer, so the design of
KLAWS-2G also followed that of KLAWS with a 15\,m mast and
temperature sensors at {10} heights from $-1$\,m to
14\,m\ (every 2\,m from 0--14\,m), and {seven} propeller
anemometers at the seven heights above 1\,m \citep{Hu19}.  There was
also one air pressure sensor and one relative humidity sensor at
2\,m height.
A data acquisition electronic box was placed at the foot of the mast
and connected to PLATO-A, sharing the {\it MAIN} computer of AST3.  A
web page was developed to show real-time data from KLAWS-2G and
historic data could also be
visualized\footnote{\it \url{http://aag.bao.ac.cn/klsite/klaws2g.php}}.

KLAWS-2G collected data since January 2015 continuously for 20 months
until the top part of the mast above 5\,m was broken due to a loose
guy-wire and strong gusts, as was concluded during the 33rd CHINARE.

\subsubsection{CSTAR-II}
\label{sec:ncstar}

The original CSTAR was decommissioned and later modified to have pointing
capability.  Two of the four telescopes were co-mounted on a AP\,1600
equatorial mount to build the new CSTAR (CSTAR-II) with different
filters of $i$ and $R$, respectively.

Tests of CSTAR-II were also done at Mohe with AST3-2 for about 5 months
\citep{Zhu20} and it was able to run automatically through a night
once started manually at twilight.
It was installed and tested at Dome~A successfully before it was
turned off in early April due to frosting, stuck mount, and the fact
that it drew too much power compared to its small size and potential
of producing science data.


\subsection{The 32nd CHINARE}
\label{sec:32nd}

Besides servicing PLATO-A, the team focused on making AST3-2 work
again.  A lot of work {was} done on its motion control
system, CCD heat dissipation system, and defrosting blower inside
tube, etc. An external defrosting blower system was installed on a
tower in front of the telescope (Fig.~\ref{fig:site32}).
An upgraded CODS system replaced the old one for AST3-2 and AST3-1.

To solve the problem of limited power, a clean energy system was
developed at PMO and deployed to Dome~A.  It had more solar panels
and a wind turbine designed to generate power at low wind speed
(e.g., {4\,m\,s$^{-1}$}).  However, the wind turbine
did not work at all and did not help with the power issue in winter,
while the solar panels provided great power when the Sun was up.

AST3-1 telescope was not maintained and the rack of its native control
system was moved to a new small cabin with thermal insulation for
storage in order to free space inside the IM of PLATO-A.  The cabin was
originally planned to get power from PLATO-A and run AST3-1, but the
plan was abandoned after the situation of AST3-1 and power budget was
evaluated.

Efforts had been made to fix the problems of defrosting and stuck
mount of CSTAR-II.  It was tested to work well before the team left,
but the mount stopped moving later when temperature dropped.

KLAWS-2G had been working for a full year.  An air pressure sensor at
2\,m and an anemometer at 4\,m were replaced to restore its full
power.

Some more webcams were installed and turned out to be very useful in
assisting AST3-2 operation in winter.
Lots of engineering works and tests were carried out remotely before
twilight.  Later, AST3-2 was able to observe with occasional problems
of getting stuck.  The CODS system ran the SN survey fully
automatically with no need of intervention by a remote observer unless the
survey was interrupted by instrument problems.  The exoplanet search
was allocated the best dark time of polar nights for non-stop
short-period transit search which was run with semi-automatic scripts
supported by CODS software.  Lots of scientific results were
obtained from this season in 2016 (Sect.~\ref{sec:exoplanet}).

AST3-2 had been collecting data until {2016
June 23} when there was a high wind above
{8\,m\,s$^{-1}$}. The operation was stopped to
protect the telescope, which however could not move again likely due
to snow/ice accumulation.


Besides the astronomical activities at Dome~A, a telescope was also
installed at Zhongshan Station for winterover.  The Bright Star Survey
Telescope (BSST) is optimally designed for searching planetary
transits of bright stars \citep{Lizy15,Tian16,Zhang16}.
The telescope has an aperture size of 30\,cm with a 4k$\times$4k CCD
camera, resulting in an FOV of 3.4\degr$\times$3.4\degr.  It is able
to run robotically, and the winterover personnel at Zhongshan Station
could help to fix small problems if there was any.   BSST operated for
one season in 2016 before being brought back.

\subsection{The 33rd CHINARE}
\label{sec:33rd}

PLATO-A was able to run through the winter on engines and then mostly
on solar power until the traverse team arrived.  The focus of this
team was still to revive AST3-2 and service PLATO-A.  It was confirmed
that the gears of AST3-2 were jammed by ice especially for the RA axis
and more protections were added to prevent this from happening again.
The CODS system was upgraded and valuable data from last year
retrieved.
Following this service mission, AST3-2 operated through the whole
observing season for the first time \citep{Lixy18}.

Since AST3-1 had been at Dome~A for 5 years and problems could not be
solved on-site, it needed a thorough maintenance.  So decisions were
made for the team to disassemble AST3-1, carefully pack the parts
including optics, and bring them back.

KLAWS-2G stopped working shortly after the top part of its mast above
5\,m broken in August 2016 after collecting data for 20 months.  The
team concluded that the mast was damaged by strong gusts after a
guy-wire broke and failed to stabilize the mast.
It was then maintained and the sensors below 5\,m resumed to work for
another 17 months when PLATO-A ran out of fuel.

\subsubsection{KLCAM}
\label{sec:klcam}

KunLun Cloud and Aurora Monitor (KLCAM) is an all-sky camera similar
to HRCAM (Sect.~\ref{sec:hrcam}) or HRCAM2 that was decommissioned the
year before.  KLCAM was developed because long-term monitoring of the
sky for cloud cover and aurora contamination is crucial for site
testing at Dome~A.
KLCAM has different but enhanced thermal control designs to cope with
the harsh winter at Dome~A \citep{Shang18}.  The optical system of
KLCAM includes a commercial Canon EOS 100D camera and a fisheye lens
(Sigma 4.5mm F2.8) to cover essentially the entire 180\degr\ sky.  A
customized ARM-based embedded computer operates the camera as well as
an active thermal control system for keeping the camera at its working
temperature above 0\degrc\ even when the ambient temperature drops
below $-$80\degrc.

KLCAM was installed on the roof of PLATO-A IM.  It collected very
useful images continuously for 17 months before power was off, and it
also helped with AST3-2 operation in winter.


\begin{figure*}
   \centering
   \includegraphics[width=7.1cm, angle=0]{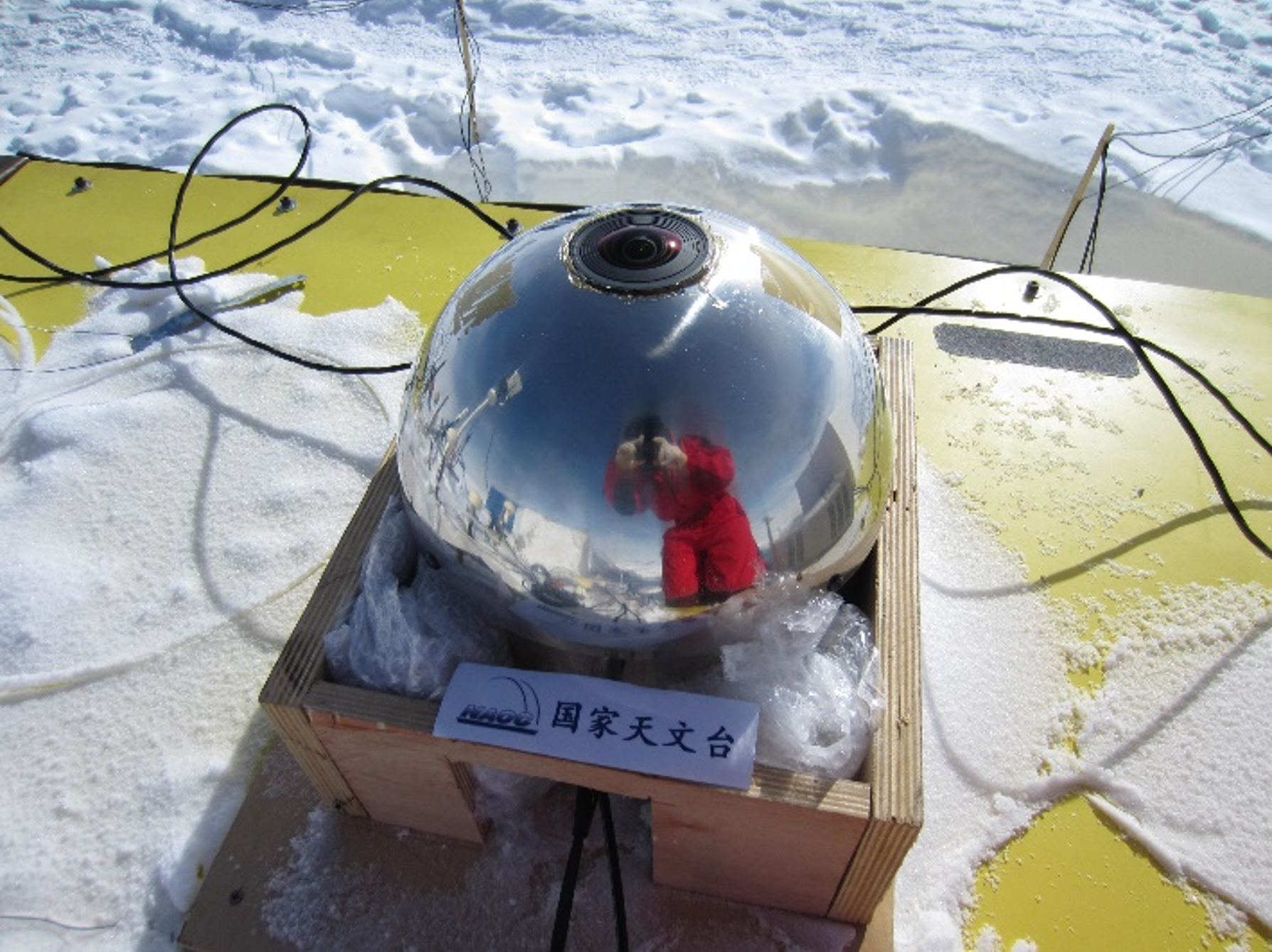}
   \includegraphics[width=5.5cm, angle=0]{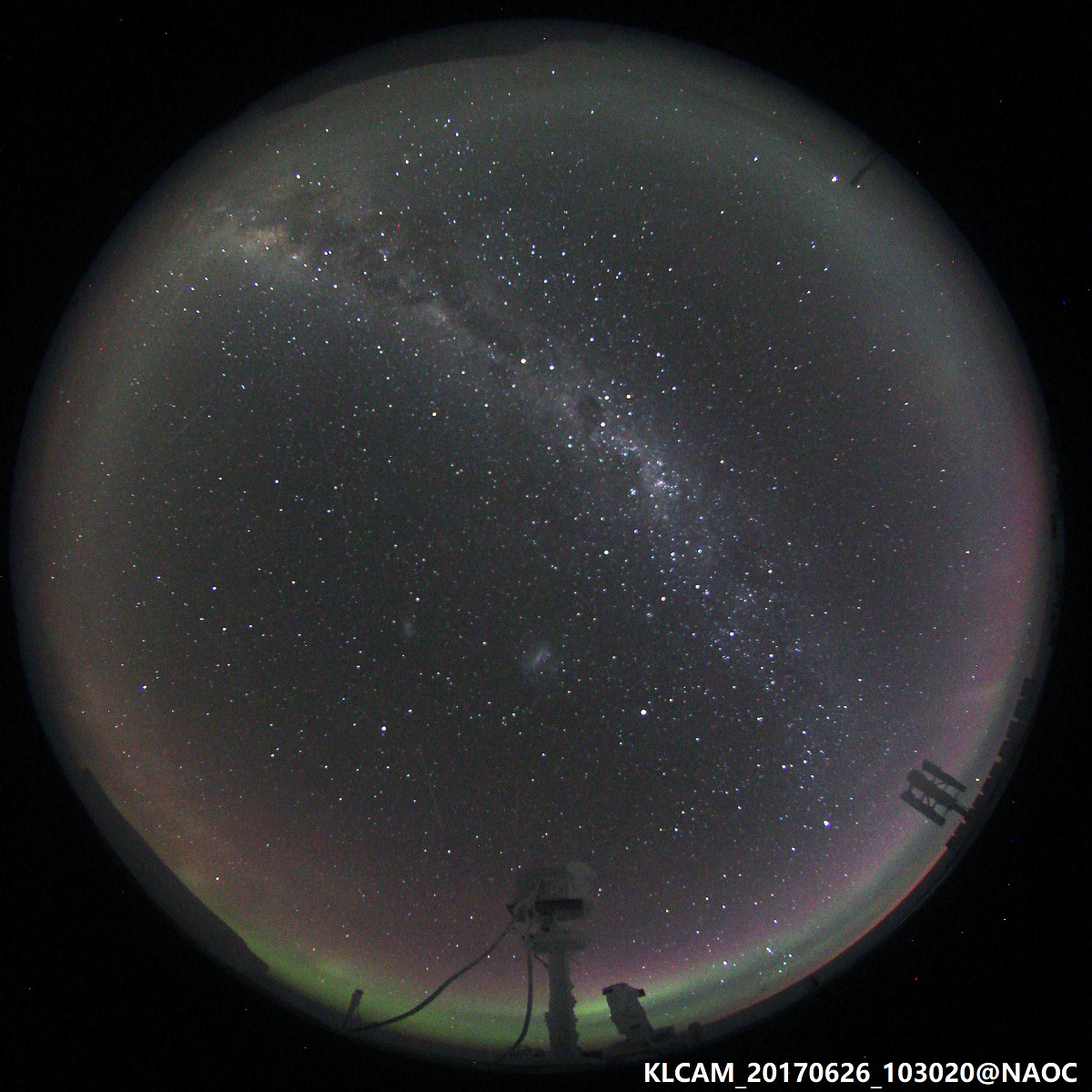}
   \caption{\baselineskip 3.8mm {\it Left}: KLCAM on the roof of PLATO-A IM.  {\it Right}: A typical
all-sky image taken by KLCAM.  The Milky-way, LMC, and SMC are
visible.  Around the horizon are aurorae in green and purple.
At the bottom are a webcam on the roof and AST3-2 in the background.
The wind turbine appears at the lower-right corner.
Courtesy of Bin Ma and Xu Yang.
}
   \label{fig:klcam}
\end{figure*}

\subsubsection{WiFi link}

A WiFi link was established from PLATO-A IM to the sleeping quarters
of the team about 400\,m away by using a parabolic antenna booster.
This made it very convenient and more effective for controlling PLATO-A
and communicating off the site for astronomy.

\subsection{The 35th CHINARE}
\label{sec:35th}

Since there was no Dome~A traverse during the 34th CHINARE, PLATO-A
struggled to operate for a second year and finally ran out of fuel
on {2018 May 27} after 17 months.  KLCAM and
KLAWS-2G also stopped collecting data then.

\begin{figure*}
   \centering
   \includegraphics[width=14.0cm, angle=0]{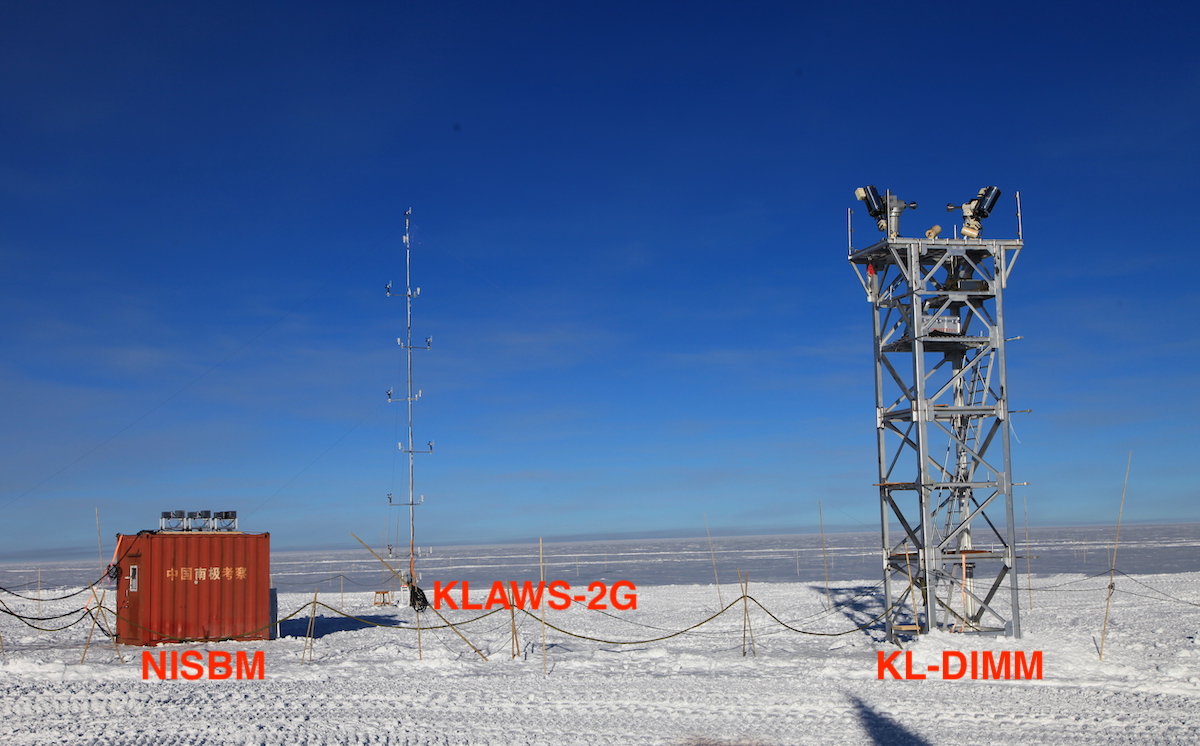}
   \caption{\baselineskip 3.8mm A photo taken in early 2019 showing two
KL-DIMM telescopes on an 8\,m tower,
the 15\,m tall KLAWS-2G, and the NISBM installed on top of
a modified shipping container that was used as a
working cabin during the 35th CHINARE. }
   \label{fig:site35}
\end{figure*}

Although KLCAM was intact, two new redundant KLCAMs were deployed and
replaced the old one to guarantee continuous long-term data collection
and avoid any possible problems developed in the old one, such as the
shutter lifetime.
A new 15\,m tall KLAWS-2G was set up and replaced the damaged one
of only a 5\,m tall mast remained, restoring its full power of
profiling the boundary layer and vertical temperature inversion
\citep{Hu19}.

PLATO-A was serviced with new engines and batteries, etc. and
revived. Lots of other work {was} done on supervisor
computers, power distribution boxes, iridium system, solar panels,
etc., either replacing devices or fixing problems on-site.

AST3-2 was also maintained with new cabling, mirror defrosting system,
and protection mechanisms preventing gears from icing, etc., making it
fully functional again.
However, as the Sun set, AST3-2 started to suffer from problems of
moving difficultly possibly because of snow and ice accumulation
again.  Some engineering work {was} done as well as
some experiments, such as using very short exposures of star pairs
to measure seeing, a new technology developed with the AST3 project
\citep{Hickson19}.

PLATO-A suffered from communication problems from time to time,
likely due to a conflict in local network setting because there were
many new instruments installed.  This seemed to be resolved after
turning off the instrument.  On {2019 August
3}, followed that PLATO-A stopped communicating, a power outage
occurred. Instruments that were still working thus stopped
observing.
In addition, PLATO-A was unfortunately not able to recover as expected
with solar power when the Sun was up.  Moreover, there was no Dome~A
traverse during the 36th CHINARE in the 2019/2020 season.  A service
mission is expected to return to Dome~A for the 2020/2021 season by
the 37th CHINARE.

\subsubsection{KL-DIMM}
\label{sec:kldimm}

Astronomical seeing is one of the crucial parameters in evaluating a
site for optical/IR observations.  Attempts had been made during the
27th and 28th CHINARE and daytime seeing measurements were obtained
for 3 days (Sect.~\ref{sec:seeing}).  The KunLun Differential Image
Motion Monitor (KL-DIMM) was therefore developed to measure and
monitor the night-time seeing for a long period so as to obtain
convincing statistics and a conclusive assessment.

It is hard to operate a DIMM telescope at Dome~A because it requires
pointing and tracking a bright star, but its FOV is usually very
small, around 10\arcmin.  Furthermore, it also has to
cope with the adverse conditions that AST3 faces.  On the other hand,
lessons learned and experience gained from CSTAR, AST3, and
CSTAR-II have helped the development of KL-DIMM.

KL-DIMM has an equatorial mount, a 10-inch (25\,cm) telescope, and a
customized mask of two subapertures attached with a 5\,cm prism each.
The GSO 10\arcsec\ f/8 R-C telescope has a carbon fibre tube which is
light and more importantly has a much smaller thermal expansion
coefficient than metal ones, minimizing focus change.  An AP\,1600GTO
mount was chosen as it had been used for CSTAR-II and at Dome~C.
However, lots of modifications on its electronics and mechanical parts
were made for automated operation in low temperatures and with
problems of icing/frosting in winter \citep{Ma18b}.

The system also has a finderscope with $f=50$\,mm and a focal ratio of
f/1.4.  Two industrial IMPERX cameras with interline CCDs of
$1600 \times 1200$ pixels
and
$3296 \times 2472$ pixels
are equipped for the DIMM telescope and
the finderscope, respectively.  This results in an FOV of about $20\degr \times
15\degr$ for the finderscope and about $15\arcmin \times 11\arcmin$
for the DIMM.

To avoid ground-layer turbulence, a customized 8\,m tall tower was
set up with another 1.5\,m base under snow surface.  One of the key
requirements for the tower design was that the jitters of the top
platform cannot exceed 1\arcmin\ under a
{8\,m\,s$^{-1}$} wind in order to better keep a star
inside KL-DIMM's small FOV.

Dedicated software was developed for automating the operation of
KL-DIMM, including observation planning, a pointing model built with
{\sc tpoint}, and data reduction, etc.  All raw image data were saved
and backed up on disks while seeing measurements were transferred back
in real-time.

For overall success, two KL-DIMMs with identical optics but different
thermal protections and mechanical modifications were deployed,
installed on the tower and started to operate during the daytime.  They
worked fully automatically, staring at Canopus all the time, as
designed after the team left Dome~A.
KL-DIMM1 had some problems of defrosting and images out-of-focus,
while KL-DIMM2 worked nearly perfectly through winter, collecting
seeing measurements in polar nights for the first time and
demonstrated the superb seeing at Dome~A
\citep[][Sect.~\ref{sec:seeing}]{Ma20a}.
KL-DIMM stopped working in August 2019 when communication and power
were lost.

\subsubsection{NISBM}

The near-infrared sky brightness monitor (NISBM) used the InGaAs
photoelectric diodes as detectors for three bands of
{$J$}, {$H$}, and {$K_s$}
\citep{Zhangj19}. Each band had an independent unit with
{$K_s$} band having an extra blackbody calibrator. They
were successfully installed and operated for three months including
about one month with nighttime (Fig.~\ref{fig:site35}). Data
analysis is in progress \citep{Tang20}.

\subsubsection{MARST}
\label{sec:smct}

The Multi-band AntaRctic Solar photometric Telescope (MARST) was
designed to have two telescope tubes co-mounted on an AP\,3600 mount
for observing solar photosphere and chromosphere at the same time,
respectively.  The telescope aperture is 15\,cm for photosphere and
13\,cm for chromosphere.  Both telescopes have an FOV of
40\arcmin$\times$40\arcmin, covering the entire solar disc (Haiping Lu \&
Peng Jiang, {\it private communications}).  MARST was installed and
operated successfully during the daytime.

\begin{figure*}
   \centering
   \includegraphics[width=10.0cm, angle=0]{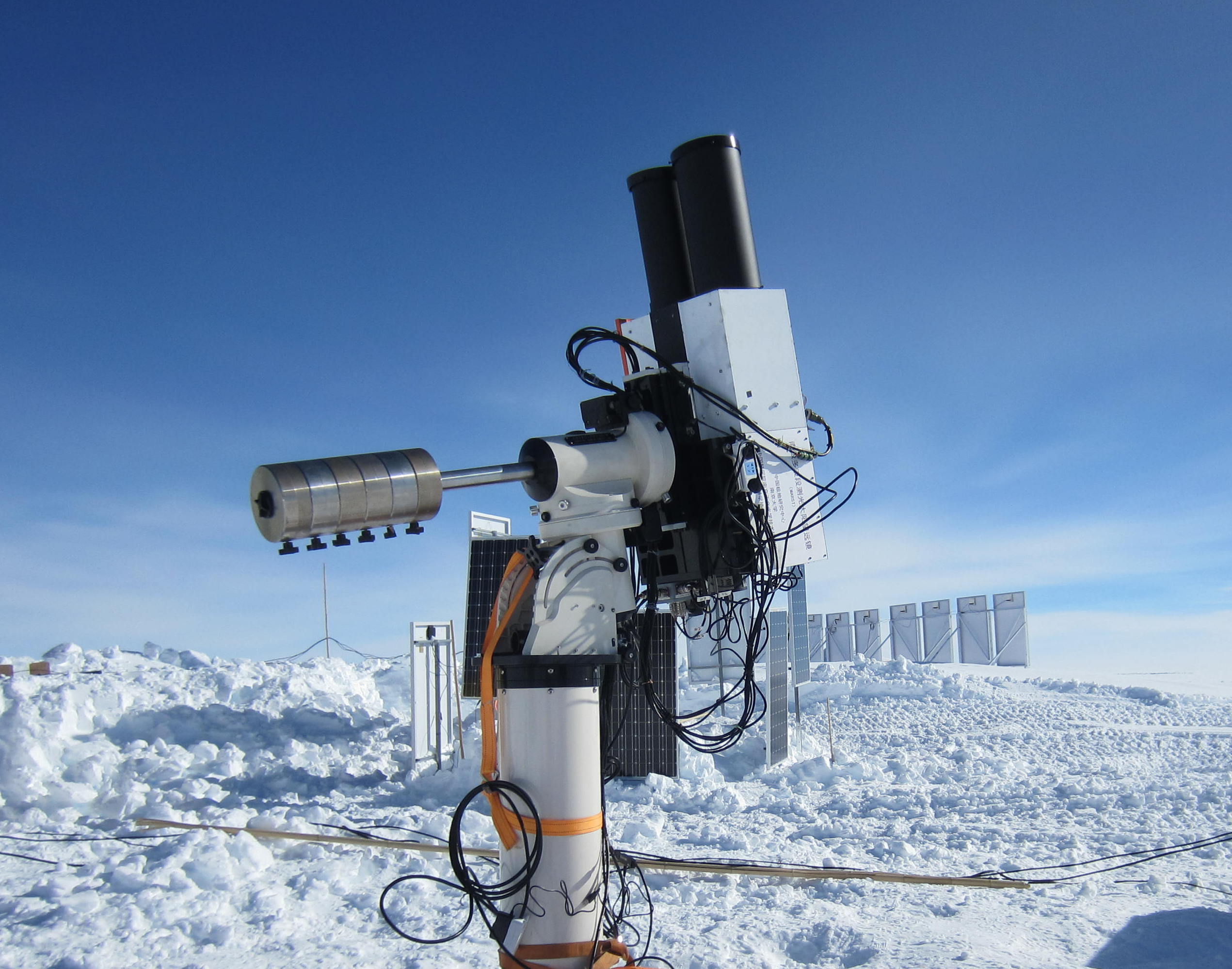}

   \begin{minipage}{4cm}
   \caption{MARST at Dome~A.}
   \label{fig:marst}\end{minipage}
\end{figure*}

Moreover, daytime observation{s} of stars were also carried out.
Benefited from the extremely clean air and clear sky, late type
stars as faint as 5 mag were observed in visual band during the
daytime.  This demonstrated the advantages and great potential of
daytime astronomy at Dome~A.
MARST was left at Dome~A for operation again by the next traverse
team.

\subsubsection{Microthermal experiment}

Another daytime experiment was to measure optical
turbulence of near-ground atmosphere.  Five microthermal sensors
were attached to the 8\,m KL-DIMM tower at different heights above
1.5\,m and the vulnerable sensors could be replaced by the team if
needed.  A sonic anemometer was installed at about 6\,m height.
The experiment collected data for about 20 days and was decommissioned
before the team left.


\section{Astronomical Research}
\label{sec:science}

Both CSTAR and AST3 were designed for imaging observations of
time-domain monitoring to take advantage of the uninterrupted dark time
in austral winter, although CSTAR was also the first Chinese
instrument for site testing \citep[][Sect.~\ref{sec:cstar}]{Zhou13}.

Figure~\ref{fig:sunmoon} illustrates the available dark time at Dome~A
in 2020.  Astronomical twilight
is defined as when the elevation angle of the Sun \sunangle\ is
between 12\degr\ and 18\degr\ below horizon.  If taking this
definition, there are {1677} hours for astronomical
night/darkness between dusk when twilight ends and dawn when
twilight starts ($\sunangle<-18$\degr), but no continuous dark time
is available. In fact, astronomical observations can usually start
even when the Sun is at about $-$10\degr\ with somewhat high sky
background.  This almost doubles the observing time to
{3112} hours and gains 68 day continuous dark time.
For more strict requirements, a new astronomical dusk and dawn can
be defined when the sky background becomes minimal and stable.
Benefiting from the extremely clean air, this happens when the Sun
is below $-$13\degr\ at Dome~A
\citep[][Sect.~\ref{sec:skybackground}]{Zou10}, providing
{2605} hour observing time with 32 day uninterrupted
dark time.

It should be pointed out that even with the new definition of
astronomical dawn/dusk at Dome~A, the observing time of
{a} year is still less than the total observing time by
regular definition of astronomical twilight/night
($\sunangle<-18$\degr) at Mauna Kea, North Chile, and La Palma which
all have more than {3300} hour dark time.

\begin{figure*}
   \centering
   \includegraphics[width=14.0cm, angle=0]{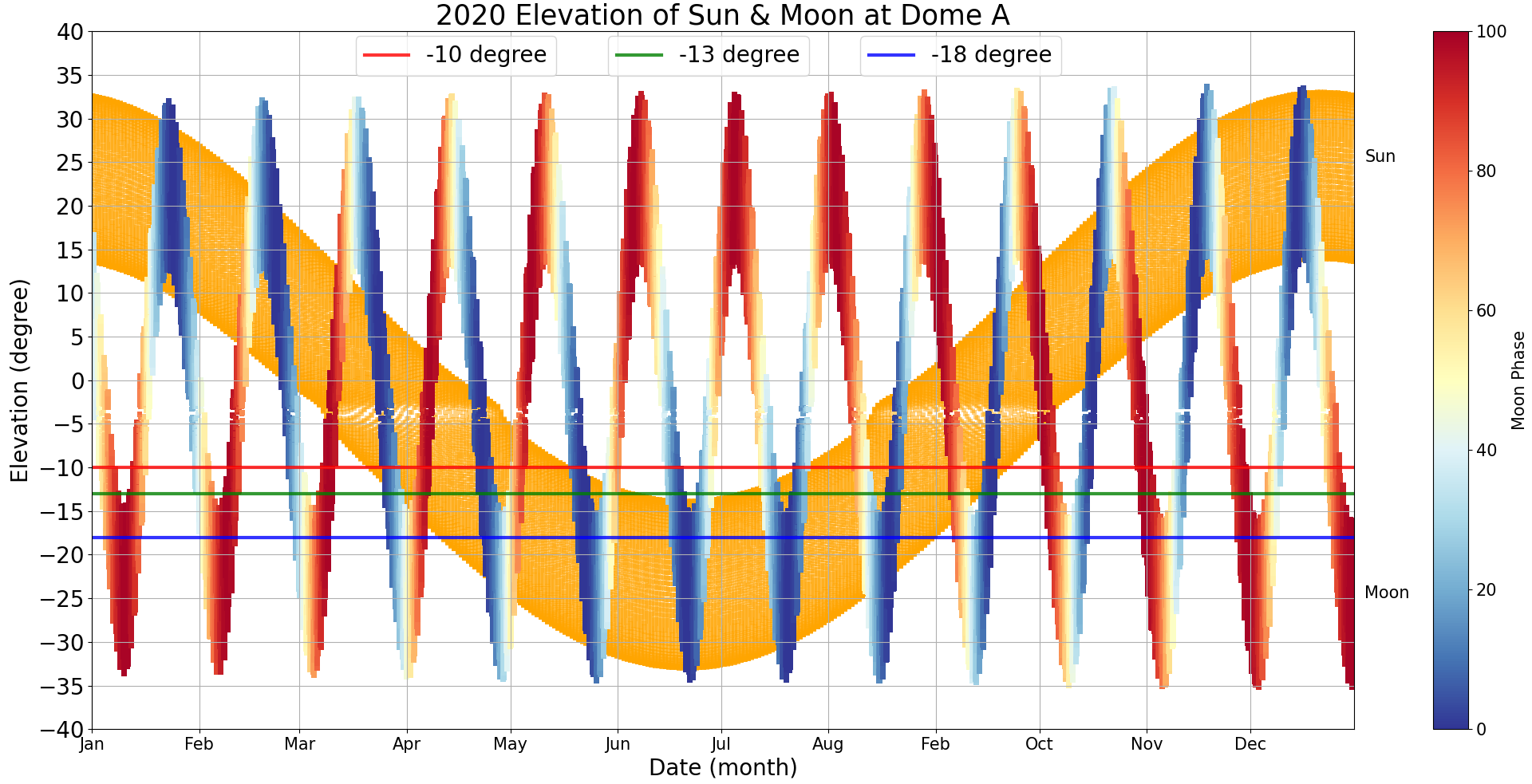}
   \caption{\baselineskip 3.8mm The elevation angles of the Sun and the Moon in 2020 at Dome~A,
showing available dark time.  The three horizontal lines indicate the
Sun's elevation at $-10\degr$, $-13\degr$, and $-18\degr$.
The phase of the Moon is also shown using the color scale.
Courtesy of Xu Yang.
}
   \label{fig:sunmoon}
\end{figure*}


\subsection{Photometry and Data Products}
\label{sec:photo}

\subsubsection{CSTAR data products}
\label{sec:cstardata}

The data obtained with CSTAR were unprecedented, as the telescopes
stared at an area of about 20\,deg$^2$ centered around the south
celestial pole for the entire observing season every year, taking an
image every 20 or 30\,s.  As CSTAR did not track, the sky
area rotated on the CCD images one cycle every day, but most of the
stars were the same in the images.  Due to technical problems, the
most useful data are $i$-band data in 2008 and 2010, as well as
defocused $g$ and $r$ band data in 2009.  The well-focused $i$-band
telescope did not take data in 2009 for some technical reasons.

The first catalog from 2008 data was released for $i$-band with a
limiting magnitude down to about {16$^{\rm th}$}
mag \citep{Zhou10b}. There are usually {10\,000}
stars on each of the {300\,000} images.
The images were corrected with a `super' bias frame created during
testing observations \citep{Zhou10a}, and a `superflat' from science
images with high sky background.  There was also a special `residual
flat-field' correction that used stars as illuminating sources as
they moved across different pixels during their daily circular
motion on images.  Aperture photometry was performed and the error
is 0.1 mag with $S/N=10$ at {13$^{\rm th}$}
magnitude.  More than {10\,000} stars were detected
in the field and most of them have more than
{300\,000} measurements in the 2008 season.

To further improve the photometric precision, a series of follow-up
corrections of instrument effects have been studied and carried out
using the catalog data, including ghost image correction
\citep{Meng13}, inhomogeneous extinction (cloud) correction
\citep{Wangsh12}, and diurnal effects from imperfect flat-fielding
\citep{Wangsh14a}.  The photometric precision has been improved to
about 4\,mmag at $i=7.5$ mag and about 20\,mmag at $i=12$ mag
\citep{Wangsh14b}.  A review on these efforts can be found in
\citet{Zhou13}.
The latest CSTAR $i$-band catalog and light curves are available
online \footnote{\it \url{https://nadc.china-vo.org/data/data/cstar/f}}.

Images from the other three telescopes with filters $g$, $r$, and
$clear$ have defocused PSFs.  Efforts were made to employ difference
image analysis on {800\,000} useful images of 2009,
reaching a limiting magnitude of about 13.5 mag for both $g$ and $r$
bands \citep{Oelkers15}.

Photometry on the 2010 data were also performed and light curves were
used to identify variable sources \citep{Wanglz13}.
Combining 2009 and 2010 data, \citep{Oelkers16} carried out an
in-depth search for transients, stellar flares, and variables.  More
importantly, through rigorous inspections, they identified many
systematic effects that could have resulted in erroneous claims in
some analyses using the same dataset.


\subsubsection{AST3 data products}
\label{sec:ast3data}

AST3-1 obtained useful science data in 2012, and AST3-2 in 2016 and
2017.  Both were operated in $i$-band.  General catalog from 2012 data
has been published \citep{Ma18a}
\footnote{\it \url{http://explore.china-vo.org/data/ast3-survey/f}} and
photometry of the exoplanet search fields, observed during the best
polar nights in 2016, has been released \citep{Zhang19a}
\footnote{\it \url{http://casdc.china-vo.org/archive/ast3/II/dr1/}}.  Other data
from 2016 and 2017 are being released soon \citep{Yang20}.

Given some special difficulties with AST3 operating at Dome~A, some
new methods were developed for flat-fielding and dark corrections.
Twilight flats were studied and taken at optimal conditions at Dome~A,
towards anti-sun direction with altitude about 75\degr\ \citep{Wei14}.
However, the sky is still not flat, with a gradient of 1\%  across the
large FOV of AST3.  The slope was fitted and removed from each flat
image before combining them into a master flat-field image.

Due to heat dissipation problems, the CCD camera was often operated
at temperatures between $-40$ to $-50$\degrc, resulting in a high dark
current.
Since AST3 cameras did not have a shutter, dark frames could not be
taken on-site while lab dark images had different patterns.
A new method was developed to derive a dark frame from science
images and it turned out to be efficient and greatly improved the
photometric precision \citep{Ma14b,Ma18a}.

Electrical cross-talk from multi-channel readouts appeared on images
when there were saturated stars.  It was studied and corrected for
better photometry \citep{Ma14b,Zhang19a}.
Obvious diagonal stripes on raw images due to interference from
telescope electronics were also observed in 2016 and a method
targeting the origin of the problem was developed to remove them well
\citep{Ma20b}.

The commissioning of AST3-1 in 2012 surveyed about 2000\,deg$^2$ sky
area for SN as well as the LMC and SMC, and monitored a dozen fields
including the LMC center, a testing field for exoplanet transits, and
some Wolf-Rayet stars.
For a typical 60\,s exposure, the $5\sigma$ limiting
magnitude is $i=18.7$ with a typical FWHM of 3.7 arcsec.  The
observed $1\sigma$ astrometric precision is
{0.1\arcsec} in both RA and Dec while the
internal precision can be much better for bright stars.
Flux calibration was based on the AAVSO Photometric All-Sky Survey
(APASS)\footnote{\it \url{https://www.aavso.org/apass}} DR9 catalog
\citep{Henden16}.

The data release consists of {14\,000} scientific
images, 16 million sources bright{er} than $i=19$ and 2 million
light curves. However, most observed fields did not have many
repeated observations, so most light curves are not very useful
\citep{Ma18a}.

In the 2016 season, AST3-2 covered some SN survey fields from 2012,
but the main focus was the exoplanet project named the CHinese
Exoplanet Searching Program from Antarctica (CHESPA). The program
selected target fields that would be later scanned by {\it TESS} and
was allocated polar nights in May and June for continuous monitoring
of short-period exoplanets \citep{Zhang19a}.
A group of 10 adjacent fields was planned and scanned for 37 nights
with some interruptions caused by instruments, weather, and other
programs.  This resulted in an overall time coverage of about 40\%
while it was improved to about 80\% in 2017 when a group of 22
fields was scanned.  The rotating monitoring takes three
10\,s exposures of each field each time, and the overall
cadence was about 12 minutes.

The FWHM of stars in images has a wide range with a median value of
about 4.5\arcsec\ although AST3 has a pixel scale of
{1\arcsec~pixel$^{-1}$}. This was mostly due
to the tube seeing caused by heat from the defrosting blower
(Sec{t.}~\ref{sec:ast32}) and CCD camera inside the telescope
tube. The large FWHM was actually good for exoplanet search
observing bright stars, but the limiting magnitude could not go deep
for other studies of faint sources, such as SN search.

More than {35\,000} images were take{n} for
CHESPA in 2016 and from this dataset more than
{26\,000} light curves of stars bright{er} than
$i=15$ mag were obtained and released with and without detrending
and binning to a cadence of 12 minutes. The best photometric
precision at the optimum magnitude around 10\,mag is around 2\,mmag.

\subsection{Exoplanet}
\label{sec:exoplanet}

The continuous dark time at Dome~A and the short cadence design of
CHESPA are sensitive for searching short-period transiting exoplanets.
The scanning fields were chosen specifically within the Southern
Continuous Viewing Zone (CVZ) of TESS, so as to maximize follow-up
observations and the chance of detailed studies of their atmosphere
and internal structure.

More than 200 objects with plausible transit signals were detected
from the 2016 AST3-2 data and 116 of them were classified as candidates of
transiting exoplanets based on their stellar properties and reasonable
planetary radii derived \citep{Zhang19b}.

Figure~\ref{fig:transient} shows an example light curve of one candidate.
Detailed fitting to the transit signals has revealed their parameters,
including period, transit depth, and duration, etc.  The host stars
range from $i=$7.5\,mag to 15\,mag and the typical transit depth are
20--50\,mmag.
The orbital periods of the candidates are from 0.2 to 6 days.
Follow-up confirmation with radial velocity observations are in
progress.
Using the same dataset, methods involving machine learning and image
subtraction to detect transients were studied \citep{Huangtj20}.

\begin{figure*}
   \centering
   \includegraphics[width=14.0cm, angle=0]{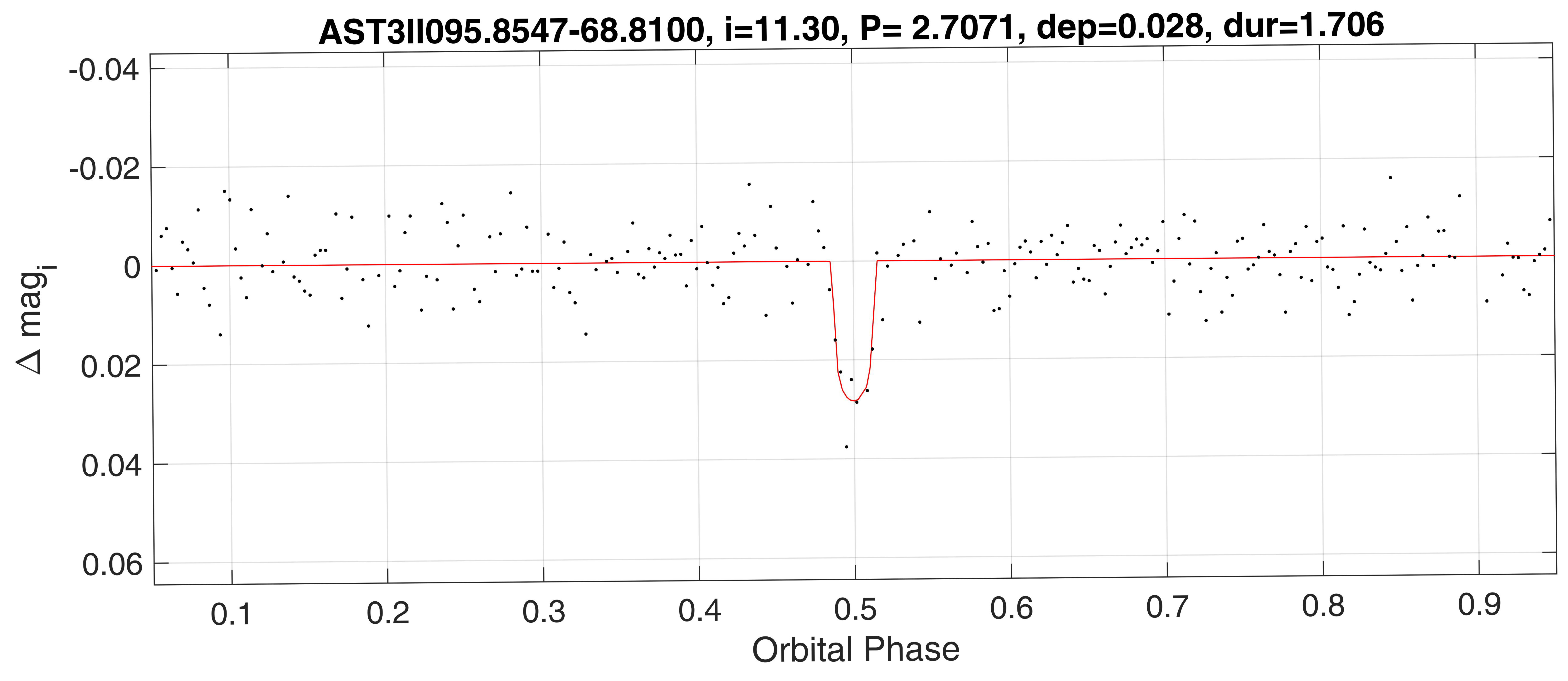}
   \caption{\baselineskip 3.8mm The phase-folded light curve of an exoplanet candidate.
Shown on the top are its AST3 name, $i$ magnitude, period in days,
transit depth in mag, and duration in hours.
Courtesy of Hui Zhang.
}
   \label{fig:transient}
\end{figure*}

In addition to AST3, data mining through the $i$ band CSTAR catalog
was also productive.  Although CSTAR was not specifically optimized
for transit search, its observing mode and the improved photometric
precision of the catalog made it suitable for this purpose.

In 2008, CSTAR stared at an area of 20\,deg$^2$ centered at the
south celestial pole for more than 100 days, taking an exposure
every 20 or 30 s.
The photometric precision of light curves for 20\,s
exposures is $\sim$4\,mmag for the brightest stars of $i=7.5$ and
$\sim$20\,mmag for $i=12$.  The transit search limited to sources of
$7.5 < i < 14$. After applying very strict criteria of selection and
false-positive ejection, \citet{Wangsh14b} found 10 candidates of
exoplanets. Follow-up spectroscopic observations of the candidates
were carried out with the Wide Field Spectrograph (WiFeS) on the
Australian National University 2.3\,m telescope and
{four} of them were ruled out because of their nature of
giants.  For the remaining {six} dwarf candidates, RV
measurements were obtained with WiFeS.  None of the candidates
showed RV variation beyond the intrinsic uncertainty of 2\kms\ of
the instrument, indicating that they are not unblended eclipsing
binaries. Therefore, higher resolution RV observations are
need{ed} for confirmation.

At Zhongshan Station in 2016, BSST monitored Proxima Centauri for over
10 nights shortly after the discovery of the Earth-mass planet Proxima
Centauri b which has an orbital period of about 11.2 days.
\citet{Liuhg18} proposed to search for transit signals of the planet
and if confirmed, its atmospheric properties and habitability, etc. can
be well studied.  They reported a tentative transit event at a
confidence level of 2.5$\sigma$ and expected more high-cadence
observations to be done.

\subsection{Stellar Variability}

\subsubsection{Variable stars}
\label{variable}

For synoptic surveys like CSTAR and AST3, variable stars are always
expected to be detected.
Two groups worked on the CSTAR data independently.  Without using
the published CSTAR catalog \citep{Zhou10b}, \citet{Wanglz11}
carried out a time-series photometry on the same dataset of 2008 and
reached a similar photometric precision (Sect.~\ref{sec:cstardata}).
Over 70\% of their bright-star sample has more than
{20\,000} photometric measurements, and among many
sources of error, the internal statistical uncertainty only is less
than 1\,mmag for stars with $i\leq13.5$\,mag.  They discovered 157
new variable stars with many kinds (Fig.~\ref{fig:cstarvariable}).
The group did the same analysis of more than {9000}
stars down to $i\leq 15.3$\,mag in the 2010 data \citep{Wanglz13}.
They detected 188 variable stars, including 67 new ones, but some in
the previous studies were not recovered for various reasons.

\begin{figure*}
   \centering
   \includegraphics[width=10.0cm, angle=0]{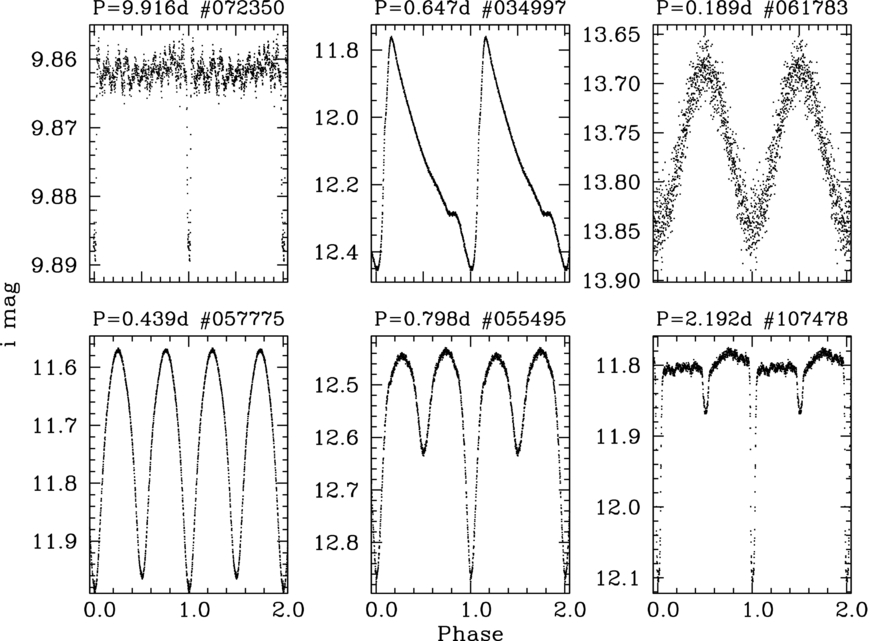}
   \caption{\baselineskip 3.8mm Examples of phased light curves of CSTAR variables
\citep{Wanglz11}.  From left to
right, {\it top}: transit source, RR Lyrae, and $\delta$\,Scuti;  {\it
bottom}: eclipsing binaries with contact, semi-detached, and detached
configurations.
Printed with permission from the authors and by permission of the AAS.
}
\label{fig:cstarvariable}
\end{figure*}

Using the CSTAR catalog of 2008 data, \citet{Wangsh15} carried out a
time-series analysis and recovered most of the variables previously
reported by another group \citep{Wanglz11,Wanglz13}.  However, there
exist some discrepancies that some variables were not confirmed while
83 new variables were reported.
\citet{Yangm15} identified and studied 53 eclipsing binaries and
presented their orbital parameters.  In addition, by studying eclipse
timing variations for semi-detached and contact systems, they
identified two interesting triple systems and derived the orbital
parameters of the third body in one system.
Another detailed study on CSTAR eclipsing binaries focused on three
individual objects \citep{Liun18}.

The defocused images from three telescopes of CSTAR ($g$, $r$, and
{\it clear}) in 2009 were also systematically analyzed (the
$i$-band telescope did not take data in 2009 for some reasons).
\citet{Oelkers15} applied the difference image analysis method
\citep{Alard98} and identified 105 variable stars, of which 37 were
not detected in previous papers.
Although not all of the stars have both $g$ and $r$ colors, and
$i$-band data were taken at different years, the color information
helped the classification and understanding of normal pulsators,
irregular variables, and eclipsing binaries in the sample.


Colors ($g$ and $i$) were also obtained with CSTAR-II during its
testing run of nearly 5 months at Mohe
(Sec{t.}~\ref{sec:ncstar}). The detection limit was about 13--14
mag.  With nearly {7000} light curves of each color
and the best overall photometric precision of about 3\,mmag for
brightest stars of 9\,mag, 63 variable stars were detected and 48 of
them were identified as new variables of eclipsing binaries or
pulsating stars \citep{Zhu20}.

CSTAR has provided the most complete dataset for studying variable
stars of periods shorter than 180 days and magnitudes down to about
$i\sim15$\,mag and $g\sim r \sim 13.5$\,mag in the sky area around the
south celestial pole.  Those who wish to use the data should refer to
all the studies as there are many differences.

AST3 has a much larger aperture than CSTAR and therefore reached a
deeper limiting magnitude despite that it was not working under the
ideal conditions due to technical problems.
From the $i$-band photometric catalog of AST3-1 2012 data
\citep{Ma18a}, \citet{Wanglz17} selected the testing field for
transiting exoplanets on the Galactic disk and carried out a
time-series analysis.  This field had the most monitoring
observations with more than {3500} images of total
38.9 hour integration time over 8 days.
There are more than {90\,000} sources down to
$i\leq16.5$\,mag in this single field and they detected 560 variable
stars, of which 339 were new detections with both eclipsing binary
stars and pulsating stars. A few variable stars show unusual
behavior, such as a plateau light curve or a secondary maximum,
etc., and need to be investigated further.

Using light curves of the same field as for the above study,
\citet{Shi19} analyzed three W~UMa-type contact binary systems and
derived their degrees of contact, mass ratios, and physical
parameters.  Some interesting results were obtained.  They found that
the orbital period of one system (AST19571) is increasing, possibly
indicating mass transfer in the system from the less massive star to
the more massive one. Another system (AST38503) is a deep-contact
binary with a degree of contact of 66\% and a mass ratio of 0.314,
which is possibly evolving into an extremely deep overcontact state and could
eventually merge into a single star with large angular momentum loss.
A pioneer work of this subject was also done using CSTAR data on one
W~UMa-type system and suggested the presence of a close-in third body
\citep{Qian14}.

Other than AST3 and CSTAR, although the main purpose is for site
testing to monitoring the sky background and extinction, the data of
wide-field cameras, such as Gattini, HRCAM, and KLCAM, can also be
used for time-serious analysis of very bright stars
\citep{Yangy17,Sims13,Yang20} and the potential is yet to be explored
more.

\subsubsection{Asteroseismology}
\label{astero}

High-precision, high-cadence time-series photometry from Dome~A are
ideal for asteroseismology study to probe the internal structure of
stars \citep{Fu14}.

Using CSTAR light curves spanning three years in three colors
($i$-band in 2008 and 2010, $g$ and $r$-bands in 2009) and having
total more than 250 periods, \citet{Huang15} carried out a detailed
analysis on a single RR Lyrae variable {\it Y Oct}.  Fourier analysis
of the data revealed only the fundamental frequency and its harmonic
frequencies in three bands, indicating the source being a non-Blazhko
RRab.  Combined with archival data, they derived a period change rate
of $-0.96\pm0.07$\,days Myr$^{-1}$, much smaller than typical RRab
stars.  The physical parameters, such as metallicity, $T_{\rm eff}$
etc., were also derived for this star.

Similarly, using more than 1950 hours of high-quality light curves in
CSTAR $g$ and $r$-bands from 2009, \citet{Zong15} discovered
low-amplitude oscillations in the star HD\,92277, which was classified
as a $\delta$\,Scuti star.  They detected 14 and 21 pulsation
frequencies in the $g$ and $r$ bands, respectively.  An accurate
period of 0.0925 days was also derived.  Multi-color observations were
shown to be valuable for mode identification.

\subsubsection{Stellar flares}

Stellar flares are common but rarely recorded, because only long-term,
uninterrupted observations can detect them systematically.  Data from
CSTAR and AST3 during Dome~A winter are therefore very valuable for
identifying and studying these events.

\citet{Qian14} presented 15 stellar flares when studying a contact
binary system in 2010 CSTAR data.  However, \citet{Oelkers16}
pointed out that this was a false detection due to the diurnal
motion of a ghost image of a bright star.  They claimed that
$\sim20\%$ of 2010 CSTAR light curves exhibit similar features that
could be mistaken as flares.  After carefully rejecting ghosting
signals and other global artifacts, they identified 29 flaring
events in 2009 and 2010 CSTAR light curves and derived a flare rate
of $7\pm1 \times 10^{-7}$ flare~h$^{-1}$ for the entire CSTAR
field, $5\pm4 \times 10^{-7}$ flare~h$^{-1}$ for late K
dwarfs, and $1\pm1 \times 10^{-6}$ flare~h$^{-1}$ for M
dwarfs in halo.

\citet{Liang16} searched for flares in the 2008 CSTAR light curves
and identified 15 flare events in 13 out of
{$>$18\,000} stars. They modeled the flares and
presented flare amplitudes between 1\% -- 27\% and durations from 10
to 40 minutes.  They also found a linear relationship between flare
decay time and total duration.
The same group also carried out a similar analysis of the 2016 AST3-2
data of exoplanet transit search in the CHESPA program, 20 stellar
flares from different sources were identified \citep{Liang20}.
They were able to model the stellar flares to obtain their properties,
including duration, amplitude, energy, and skewness, for future study.
The durations of the flares in this study range from 28 to 119
minutes, with most within an hour.

\subsection{SN and Other Transients}

Very early discovery of SNe was one of the key programs for AST3.
However, due to problems of large FWHM from tube seeing and large
extinction from frosting, the limiting magnitude of either AST3-1 or
AST3-2 was optimal for faint sources.  Nevertheless, there were SNe
detected and discovered by the AST3 automated survey.

As early as in 2014 during the field test in Mohe
(Sec~\ref{sec:ast32}), AST3-1 recorded SN~2014J in M82 \citep{Ma14c},
but did not discover it in real-time.  Shortly after that,  a new type
Ia SN was discovered with spectroscopic confirmation as SN~2014M
\citep{Ma14d}.
After being deployed to Dome~A, AST3-2 discovered another one, a Type
IIP SN~2017fbq.
The real-time pipeline of AST3 was also very powerful in detecting
other variable sources, such as dwarf novae \citep{Ma16}, minor
planets, AGNs, and variable stars, but most of these were not reported
in real-time and they are yet to be studied.

In 2017, AST3-2 contributed to the campaign of seeking and
successfully detected the first optical counterpart of the
gravitational wave source GW170817 \citep{Hul17,Andreoni17,Abbott17}.


\section{Astronomical Site Testing}
\label{sec:sitetesting}

Besides months of continuous dark time in winter, the nature of
Dome~A being a good optical, infrared, and THz observatory has long
been discussed qualitatively based on its stable atmosphere, low
temperature and low water vapor, etc.  However, quantitative
measurements must be made in order to be able to help the science
expectations, the designs of telescopes and instruments, the
construction planning, and future operations.

Years of site testing efforts have greatly advanced our knowledge
about Dome~A.
Studies have revealed a median free-atmosphere seeing of 0.31\arcsec,
a median boundary-layer height of 13.9\,m, the PWV in the 0.1--0.2\,mm
range, and darker sky background, etc.  These have demonstrated the
best observing conditions and the merits of Dome~A for ground-based
astronomical observations, as will be discussed in the following
sections.

\subsection{Meteorological Parameters and Boundary Layer}
\label{sec:weather}

Since Dome~A is covered by ice as thick as about
{3000} meters, due to radiative cooling of ice,
atmosphere often shows temperature inversion, an extreme stable
situation in which colder, heavier air stays below warmer, lighter
air (Fig.~\ref{fig:tinversion}). Coupled with low wind shear, this
can result in very weak turbulence and good seeing.

With a full-year monitoring in 2011 from KLAWS, \citet{Hu14} found a
strong temperature inversion existed for more than 70\% of the time
above snow surface, and up to 95\% above 6\,m.  The temperature
gradient {was} often greater than 10\degrc\ from 0\,m
to 10\,m.
These were further confirmed with 20 month data from KLAWS-2G in 2015
and 2016 \citep{Hu19}.

\begin{figure*}
   \centering
   \includegraphics[width=13.0cm, angle=0]{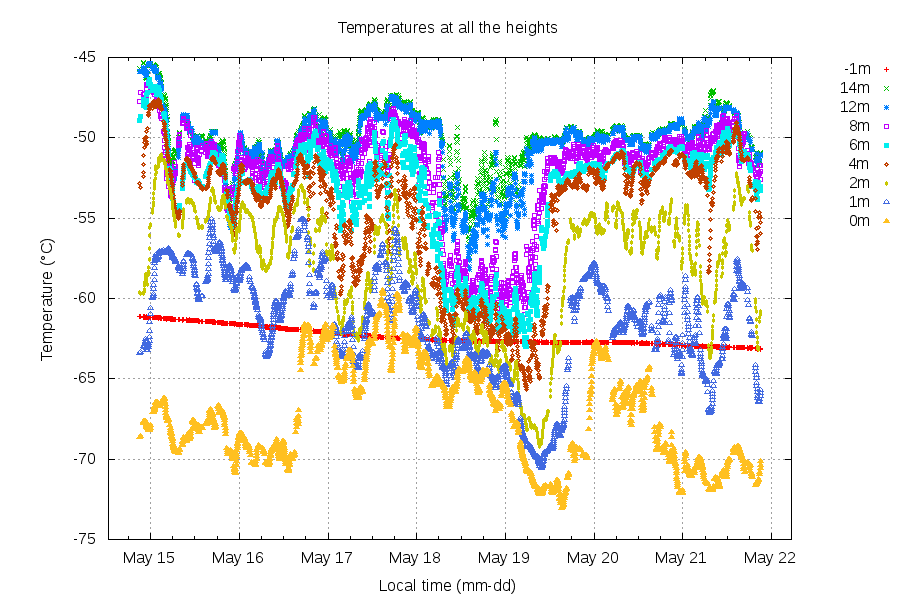}
   \caption{\baselineskip 3.8mm An example of temperature inversion in 2016 KLAWS-2G data.
Different colors indicate the temperatures at different heights.
Courtesy of Yi Hu.
}
\label{fig:tinversion}
\end{figure*}

\begin{figure*}
   \centering
   \includegraphics[width=13.0cm, angle=0]{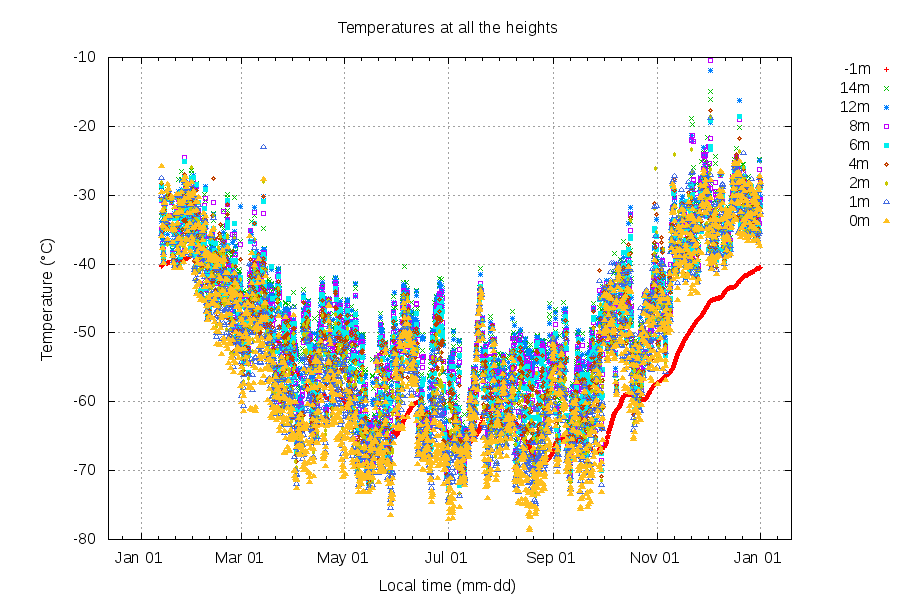}
   \caption{\baselineskip 3.8mm Plot of temperatures of 2015 from KLAWS-2G.
Different colors indicate the temperatures at different heights.
Courtesy of Yi Hu.
}
\label{fig:2015temp}
\end{figure*}

Moreover, the temperature gradient, regarded as the strength of
temperature inversion, was shown to be anti-correlated with the
thickness of the turbulent atmospheric boundary layer \citep{Hu14}.
At Dome~A, most ground turbulence exists in the boundary layer, above
which is the free-atmosphere where there is minimal energy exchange
with the earth surface and thus weak turbulence and good seeing are
expected.
Therefore, thin boundary layer and strong
temperature inversion happen at the same time, indicating
a stable atmosphere and that free-atmosphere is easy to access
with superb seeing.

The thickness, distribution and variability of the boundary layer at
Dome~A was characterized using Snodar in 2009 \citep{Bonner10}.  They
presented a median thickness of 13.9\,m, with the 25th and 75th
percentiles at 9.7\,m and 19.7\,m, respectively.   Compared to Dome~C,
the median thickness of its boundary layer was 33\,m, with the 25th
and 75th percentiles at 25\,m and 42\,m, respectively
\citep{Trinquet08}.

This has been one of the most important and exciting site testing
results for Dome~A, as it implies that not only good free-atmosphere
seeing is easy to access but the constructions of future large
facilities are also feasible.  The statistics of the boundary layer
thickness as well as the temperature inversion can help to guide the
decision of heights for telescopes and their domes, etc.

The ground temperature at Dome~A ranged from $-30$\degrc\ in summer to
about $-80$\degrc\ in winter.  During winter observing seasons, the
temperature were usually between $-50$\degrc\ and $-80$\degrc\ with
small annual variation.
Since katabatic wind dominates the Antarctica continent, the wind
speed should be very low in general with random directions at
Dome~A. These were confirmed and an average wind speed of
{1.5\,m\,s$^{-1}$} was recorded at 4\,m in 2011
\citep{Hu14}, but it was about {4\,m\,s$^{-1}$} in
2015-2016 \citep{Hu19}. They concluded that the difference reflects
the annual climatological change.


\subsection{Seeing}
\label{sec:seeing}

There are many ways to profile the vertical turbulence structures in
the atmosphere and derive seeing values, but the most direct and
commonly used instrument is DIMM \citep{Sarazin90} which measures a
total seeing above the telescope aperture.  A DIMM is usually placed
on a tower to avoid strong turbulence close to the ground.
As discussed in Section~\ref{sec:weather}, only the free-atmosphere
seeing is of great interest at Dome~A when it is measured above the
boundary layer.

Very recently, first night-time seeing as good as
{0.13\arcsec} was reported based on winter data
from KL-DIMM on an 8\,m high tower at Dome~A in 2019
\citep[][Sect.~\ref{sec:kldimm}]{Ma20a}.
The KL-DIMM took one seeing measurement every minute from April to
August with occasional interruptions for defrosting, etc.  The
measurements show a median free-atmosphere seeing of
{0.31\arcsec}, and the chance of KL-DIMM
reaching free-atmosphere was 31\% of the time at 8\,m
(Fig.~\ref{fig:seeing}, Table.~\ref{tab:seeing}).   It is estimated
to be 49\% at 14\,m based on the strong correlation between the
near-surface temperature gradient and the boundary-layer thickness.
The temperature gradient is also strongly correlated with seeing and
therefore can hopefully be used to estimate and/or derive seeing in
the future for Dome~A.

These results have demonstrated that Dome~A has the best seeing
condition on the ground.  Table~\ref{tab:seeing} compared the
night-time seeing for several good sites.  It is worth to note the
median free-atmosphere seeing and their fractions of time as these
are the really important ones for Dome~A and Dome~C.

\begin{figure*}
   \centering
   \includegraphics[width=8cm, angle=0]{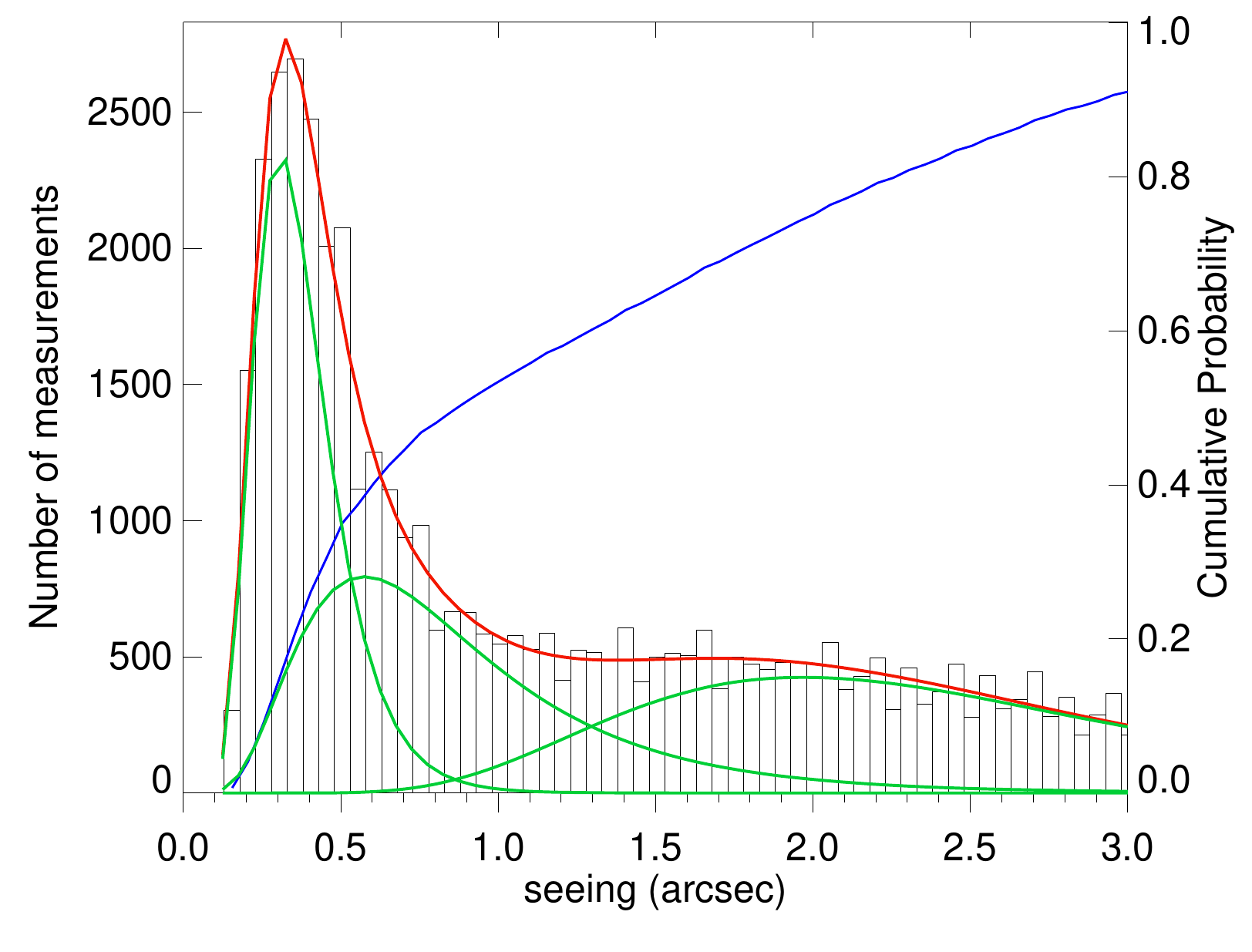}
   \caption{\baselineskip 3.8mm Distribution of night-time seeing measurements
from KL-DIMM in 2019.  The
sampling rate was one measurement per minute.  Three log-normal
components ({\it green lines}) were fitted to the data, from left to right,
corresponding to free-atmosphere,
intermediate and boundary-layer components, respectively.
The red line is the sum of the three log-normal components and the
blue line indicates the cumulative probability.
Courtesy of Bin Ma.
}
\label{fig:seeing}
\end{figure*}

\begin{table*}
\centering
\begin{minipage}[]{60mm}
\caption{Statistics of the Night-time Seeing}
\label{tab:seeing}\end{minipage}


\fns\setlength{\tabcolsep}{8pt}
\begin{tabular}{lccccccccccc}
\hline\noalign{\smallskip}
Site  & height & 25\% &  median  & 75\% & $\epsilon_{FA}$ & $\epsilon_{IN}$ & $\epsilon_{BL}$ & $f_{FA}$ & $f_{IN}$ & $f_{BL}$\\
 & m & arcsec  & arcsec & arcsec & arcsec & arcsec & arcsec & \% & \% & \% \\
\hline
Dome A & 8 &  0.41  &  0.89  &  2.02 & 0.31 & 0.57 & 1.97 & 31.0 & 30.1 & 38.9\\
Dome C & 8 &  0.83  &  1.65  &  2.32 & 0.33 & 0.54 & 1.73 & 16.2 & 14.4 & 69.4\\
Dome C & 20 &  0.43 & 0.84   &  1.55 & 0.30 & 0.42 & 1.17 & 15.7 & 29.3 & 55.0\\
Mauna Kea & 7 & 0.57 & 0.75 & 1.03 \\
Armazones & 7 & 0.50 & 0.64 & 0.86 \\
La Palma & 5 & 0.62 & 0.80 & 1.06 \\
\hline\noalign{\smallskip}
\end{tabular}

\parbox{14cm}%
{The columns are the names of sites, the heights of towers for DIMMs,
the 25th, 50th (median) and 75th percentile values of DIMM seeing, the
median values of the free-atmosphere seeing $\epsilon_{FA}$,
intermediate seeing $\epsilon_{IN}$ and BL seeing $\epsilon_{BL}$, and
their fractions of time $f_{FA}$, $f_{IN}$, and $f_{BL}$, respectively.
The entrance pupils of DIMMs are usually $\sim$ 1\,m above the top of towers.
The decomposition results for three log-normal components in seeing
histograms is only for Dome A \citep{Ma20a} and C \citep{Aristidi09}.
The seeing statistics are also shown for three best mid-latitude
sites: Mauna Kea 13N in Hawaii\citep{Skidmore09}, Cerro
Armazones in Chile\citep{Skidmore09} and La Palma in Canary
Islands\citep{Vazquez12}.
Adapted from \citet{Ma20a} with permission from the authors and
by permission of Springer Nature.}
\end{table*}

Prior to the KL-DIMM results, daytime seeing measurements of 3 days
with a DIMM mounted on AST3-1 tube in 2012 revealed
a median seeing of $\sim$0.8\arcsec\ at a height of 3.5\,m
\citep[][Sect.~\ref{sec:28th}]{Pei12}.

At Taishan Station, daytime seeing measurements of 2 days in January
2014 have shown a median value of 0.73\arcsec\ at a height of
2.5\,m above snow surface \citep{Tian20}.  As is reasonable, they also
reported a correlation between the seeing minimum and the minimum of
the atmospheric structure constant of refractive index $C^2_n$ which
was measured simultaneously using a microthermal sensor at a single
height of 2.0\,m.
As the time span is not long, it is hard to draw statistical
conclusions yet.

\cntwo\ is an important parameter denoting atmospheric turbulence.  It
is related to total integrated seeing $\epsilon$ from a height
of $h_0$ and above through

\[
\epsilon = 5.3 \lambda^{-1/5} \left (\int^{\infty}_{h_0} C^2_n(h) {\rm d} h
\right ) ^{3/5}
\]
where $\lambda$ is the wavelength of interest.  In reality, \cntwo\
can be derived from microthermal measurements.

\[
C^2_n(h) = \left (79\times 10^{-6} \frac{P(h)}{T^2(h)} \right )^2
C^2_T(h)
\]
where $P$ is the air pressure in hPa, $T$ is the air
temperature in K, and $C^2_T(h)$ is the
structure constant of temperature index which can be directly measured
with microthermal sensors at different heights.  At a certain height,

\[
D_T(r) = <[T(r_1+r) - T(r_1)]^2> = C^2_T(r) r^{2/3}
\]
where $D_T(r)$ is the structure function of temperature
and $r$ is the distance traveled by air between two sampling.


There are many ways to profile \cntwo\ vertically.  Specifically
developed for and from AST3, \citet{Hickson19} proposed the Multistar
Turbulence Monitor (MTM), a new technique using short-exposure images
of a star field.  Differential motion between any pair of star images
can be used to computer the structure function and the different
angular separations between two stars in pairs corresponds to different
heights.  The actual data analysis needs to be verified \citep{Ma20c}
and this technique can be generalized to use a small telescope at any
site.

Another method was also developed with AST3 and attempted to measure
atmospheric seeing from bright star trails on images \citep{Ma16}.
Since frame-transfer CCD was used and there was no shutter
(Sect.~\ref{sec:ast3}), during the transfer phase, a bright star will
leave a trail on the image, which is a time-series of very short
exposures (about 1\,ms) of the star.  The atmospheric turbulence can
cause jittering of the centroid of the trail and therefore can be
measured.  However, this was just to use the by-products of AST3 data
and it was coupled with telescope vibration due to wind or tracking, etc.,
further investigation are yet to be realized.

\subsection{Extinction and Cloud Cover}
\label{sec:cloud}

Cloud cover and the fractions of photometric and observable time are
important in characterizing a site for optical/IR.
\citet{Zou10} first studied the cloud cover using CSTAR $i$ band data
in 2008.  They were not able to obtain the absolute extinction not
just because the airmass did not change much in the CSTAR FOV, but
clouds, the atmosphere and frosting all contributed and could not be
easily distinguished.  Therefore they calculated relative transparency
variations from images and used their statistics to derive upper limits
of the cloud cover.  Based on excess extinction values of $<0.11$,
$0.11-0.31$, $0.31-0.75$, and $>0.75$ mag, they defined four
corresponding categories and found their fractions of time as 51\% for
little/no cloud, 23\% for thin cloud, 17\% for intermediate cloud, and
9\% for thick cloud.
They concluded that the cloud cover condition is better than that at
Mauna Kea, with a time fraction of 67\% at Dome~A compared to 50\%
at Mauna Kea (Gemini) for extinction $<0.3$ mag in $V$ band.
More details are listed in Table~\ref{tab:cstarcloud}.

\begin{table}
\centering
\caption[]{The Comparison of Cloud Cover between Dome~A and Mauna Kea
\label{tab:cstarcloud}}

\setlength{\tabcolsep}{1pt}
\fns
 \begin{tabular}{lccc}
  \hline\noalign{\smallskip}
&&\multicolumn{2}{c}{Fraction} \\
\cline{3-4}
Cloud Cover  &
Extinction ($V$)~ ~ &
Mauna Kea~ ~ &
Dome A \\
  \hline\noalign{\smallskip}
Clear       &  $<0.3$ & 50\% & 67\%\\
Patchy cloud    &  $0.3-2$ & 20\% & 31\%\\
Cloudy      &  $2-3$ & 20\% & 2\%\\
Any other usable&  $>3$ & 10\% & 0\%\\
  \noalign{\smallskip}\hline
\end{tabular}
\end{table}

The good extinction condition at Dome~A was also confirmed by others
with CSTAR data.  Based on 2008 and 2010 $i$-band data,
\citet{Wanglz13} found that the extinction due to clouds was less than
0.1 and 0.4 mag for 40\% and 70\% of the dark time, respectively.
\citet{Oelkers15} reported consistent results in 2009 $g$ and $r$ data
with the extinction due to clouds to be less than 0.1 and 0.4 mag for
40\% and 63\% of the dark time, respectively.
\citet{Wangsh12} investigated the inhomogeneous extinction due to
clouds over CSTAR's FOV of 20\,deg$^2$ and corrected the effects to
improve the photometric precision of the CSTAR catalog.  They reported
an even transparency in 83.7\% of images and concluded that more than
80\% of observation time during 2008 was suitable for accurate
photometry.

The results from CSTAR data are convincing but only representative of
the area around the south celestial pole which is about 80\degr\ above
the horizon.
However, Gattini's GASC has {an} FOV of
90\degr$\times$90\degr, also pointed near the SCP and could extend
down to 35\degr\ above horizon. Although GASC was mainly for sky
brightness monitoring in {three} colors
(Sect.~\ref{sec:skybackground}), estimation of cloud cover was also
done by following the methods and criteria for CSTAR in
\citet{Zou10}. Comparable results were obtained except that the
fraction of time for the category of the lowest extinction ($<0.11$)
dropped from 51\% for CSTAR to 34\% for GASC.

There are many factors that could complicate the estimates of cloud
cover (extinction) from images.   Frosting on the windows or entrance
pupil of instruments was probably the worst as its effects could
not be decoupled from those of real clouds on images.
Since AST3 has a much smaller FOV and suffered the most severe
frosting problem, no attempt has been tried to use its data for
cloud cover study.

Apart from studying extinction through photometry of wide-field
images, the most direct assessment of cloud cover usually uses
monitoring images produced by all-sky cameras such as ASCA for the TMT
site testing \citep{Skidmore11} or KLCAM for Dome~A.

KLCAM worked continuously for 490 days since January 2017, taking an
image every half an hour.  Visual inspection has been employed to
analyze the night-time images when the Sun was 13\degr\ below
horizon \citep{Yang20}.  Since clouds at Dome~A are always diffuse, a
semi-quantitative image classification was first defined for clouds as
clear, light, heavy, or covered.  Each image was visually checked
independently by five individuals who classified the image into one of
the four categories defined above, and the median value was taken as
the image's classification.  A confusion matrix of statistics show
very high level of agreement among the inspectors.

The results show that 83.3\% of images are classified as `clear',
11.2\% `light', 2.8\% `heavy', and 2.7\% `covered'.  The `clear'
fraction is the highest compared to other best temperate sites which
have a fraction between 70\% and 82\%.  Moreover, the `clear'
fraction are higher in May to July when the dark time is long.
Unlike CSTAR or Gattini, KLCAM was free from icing or frosting
problem owing to its special design.  Therefore, although
semi-quantitative, the results are reliable and expected.  More
quantitative work {is} necessary, such as
machine-learning analysis or photometry on the images.

\subsection{Sky Background }
\label{sec:skybackground}

The clean atmosphere with low aerosol content guarantees dark sky
background at Dome~A and it also makes the sky darken earlier.
The $i$-band sky background from 2008 CSTAR data revealed that the sky
brightness became dark and stable when the Sun was below $-13$\degr\
(Fig.~\ref{fig:twilight}), and therefore the astronomical night
(between dusk and dawn) at Dome~A can be redefined as such instead of
$-18$\degr\ \citep{Zou10}.
Using spectroscopic data from Nigel, a similar conclusion of
$-12.6$\degr\ was reached when the end of twilight was quantitatively
redefined as when the sky brightness increased by 0.5\,mag above the
median moonless dark sky \citep{Sims12a}.  They also presented the
annual dark time available as a function of wavelength between
300--850\,nm.

\begin{figure*}
   \centering
   \includegraphics[width=10.0cm, angle=0]{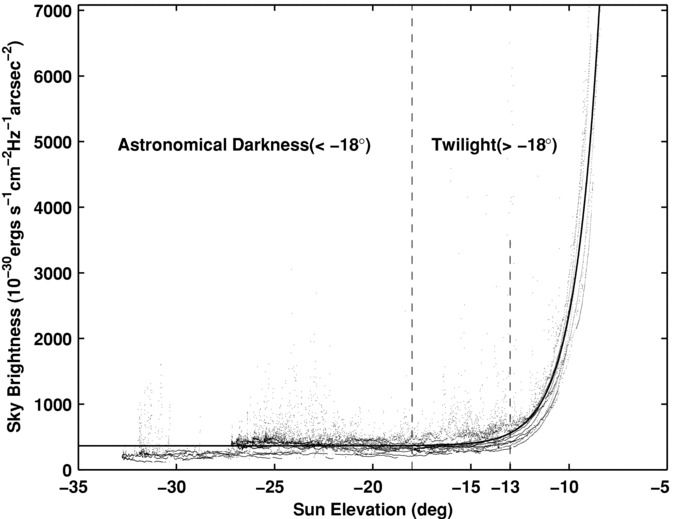}
   \caption{\baselineskip 3.8mm Relation between the sky brightness and the elevation of
the Sun, showing the sky becomes dark and stable when the Sun is below
$-13$\degr \citep{Zou10}.
Printed with permission from the authors and by permission of the AAS.
}
\label{fig:twilight}
\end{figure*}

The median $i$-band sky background of moonless clear nights was
20.5\,mag~arcsec$^{-2}$ in 2008.  This is the darkest compared to
20.10\,mag~arcsec$^{-2}$ at La Palma, 20.07 at Cerro Tololo, and 19.93
at Paranal \citep{Zou10}.
The 2010 data showed consistent results with the darkest sky
background of 20.9 \,mag~arcsec$^{-2}$ \citep{Wanglz13}.

Gattini's GASC was designed specifically for investigating optical
sky brightness over {a} large FOV of
90\degr$\times$90\degr. Measurements in three colors were made using
2009 data and the darkest 10\% in the cumulative distribution marks
sky brightness of $B=22.98$, $V=21.86$, and
$R=21.68$\,mag~arcsec$^{-2}$.  These are comparable to the values
during solar minimum at other best optical observatories in Mauna
Kea or northern Chile.

Currently, no IR sky background measurements have been published
for Dome~A.


\subsection{Aurora}
\label{sec:aurora}

One complication for astronomical site testing and observation in
Antarctica is the aurora contamination in the optical/IR bands.
Auroral emission is composed of spectral lines from atomic oxygen and
bands from molecular nitrogen and oxygen, and only a few of them are
prominent including the most common green [O\,{\sc i}] 557.7\,nm and
purple N$^+_2$ 391.4 and 427.8\,nm lines.  Customized filters can be
designed to exclude these strong auroral lines from desired bands, but
numerous weak lines still contribute to the sky brightness.
Figure~\ref{fig:auroralline} shows a useful summary of
the auroral and airglow emission lines \citep{Sims12a}.

\begin{figure*}
   \centering
   \includegraphics[width=12.0cm, angle=0]{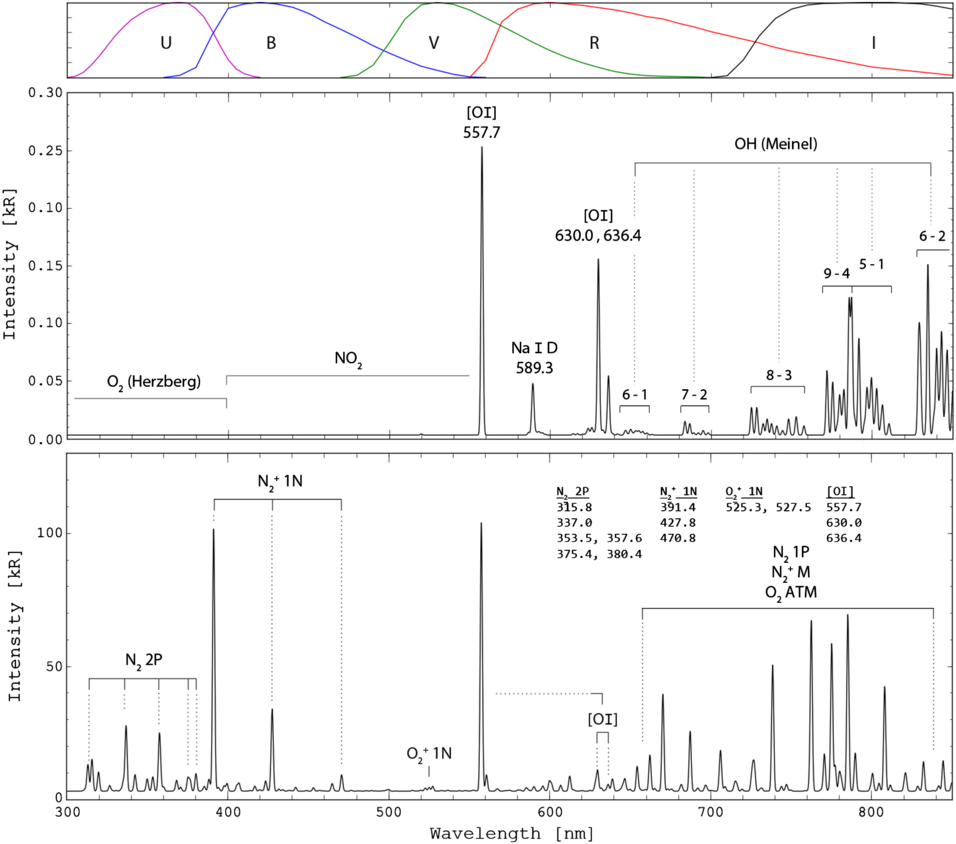}
   \caption{\baselineskip 3.8mm
Synthetic airglow ({\it middle}) and auroral ({\it bottom}) spectrum,
showing typical intensities (in kilorayleighs) of
dominant emissions in the 300--850 nm range \citep{Sims12a}.
Note that the [O\,{\sc i}] 557.7\,nm emission is hundreds times
greater in the typical aurora than in the typical airglow.
%
Printed with permission from the authors and by permission of the AAS.
}
\label{fig:auroralline}
\end{figure*}

\citet{Zou10} reported that only 2\% of the CSTAR images in 2008 were
affected by aurora.  This is certainly not representative because of
the relatively small FOV of CSTAR and the fact that aurorae usually
occur in the ``auroral oval'', a ring of 3--6\degr\ wide in latitude and
between 10--20\degr\ from the geomagnetic poles
(Fig.~\ref{fig:auroraloval}).  Fortunately,
Dome~A lies inside the ``auroral oval'', just 6\degr\ from the
South Geomagnetic Pole, therefore aurorae there generally lie below
horizon or close to horizon, leaving less contamination to the sky
areas of low airmasses.

\begin{figure*}
   \centering
   \includegraphics[width=12.0cm, angle=0]{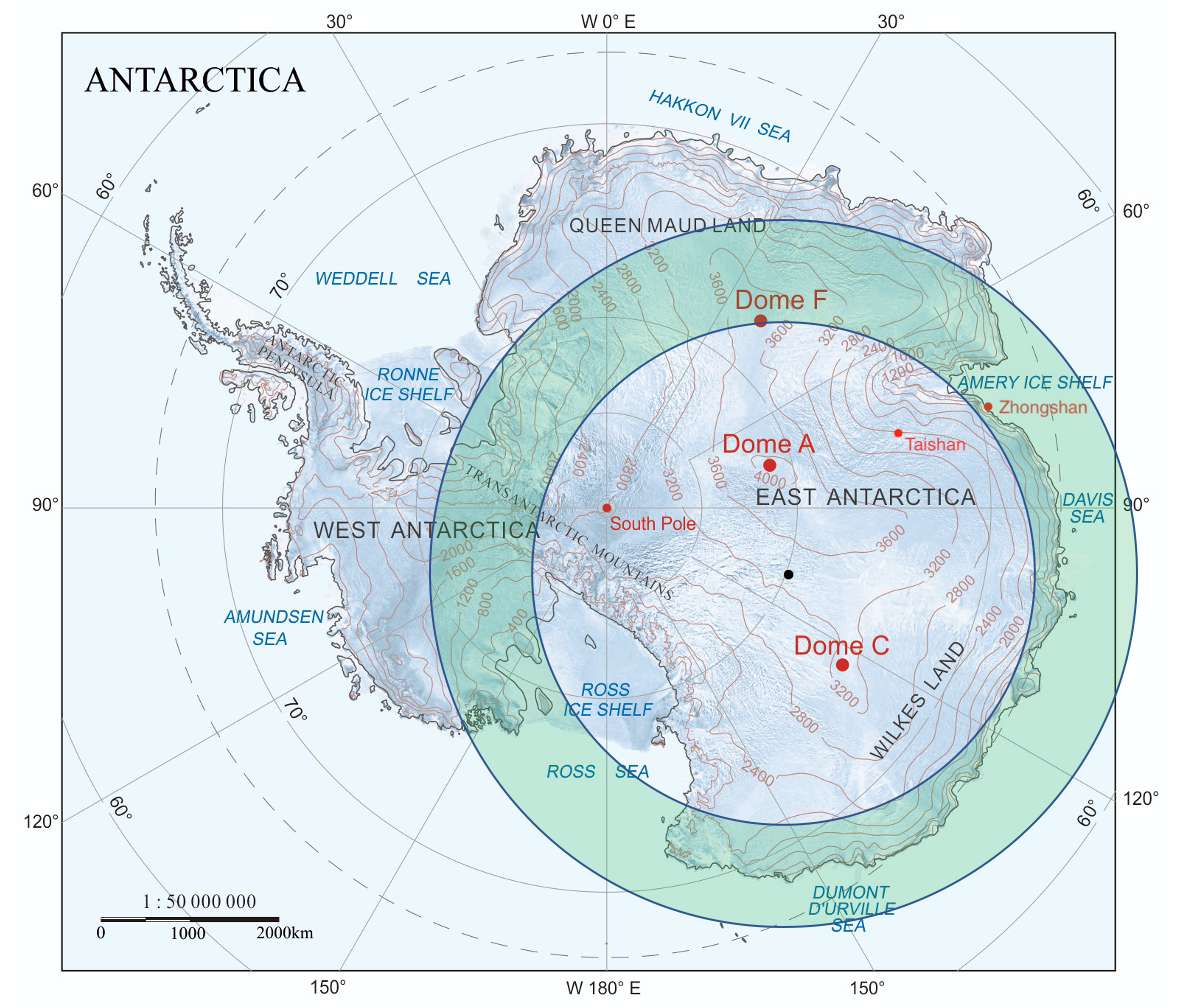}
   \caption{\baselineskip 3.8mm
Illustration of the typical location of
``auroral oval'' centered on the South Geomagnetic
Pole ({\it black dot}) in Antarctica.  It is clear that Dome~A and C are in
better situation than Dome~F and the South Pole in terms of auroral
pollution.
%
}
\label{fig:auroraloval}
\end{figure*}

The frequency of occurrence and the intensity of aurorae depend upon
the solar activity and solar cycles, but it is still important to
characterize the aurora conditions at Dome~A.
Both airglow and auroral lines were clearly detected in the spectra
from Nigel in 2009 close to solar minimum.  Having managed to separate
aurorae from airglow and calibrated the spectra with Gattini
photometry, \citet{Sims12a} studied the dominant emission lines and
found the median auroral contribution to the $B$, $V$, and $R$ bands
is 22.9, 23.4, and 23.0 mag arcsec$^{-2}$, respectively.  They also
noted that up to 50\% of observations were above general airglow
intensities, indicating aurora contamination.

Similar to the classification of clouds in KLCAM data
(Sect.~\ref{sec:cloud}), aurorae were also visually inspected by five
individuals independently on each of thousands of all-sky images taken
at night in 2017 and 2018 \citep{Yang20}.
%
Again they defined four categories semi-quantitatively according to
the intensity and extent of auroral contamination and found that 55.3\%
of the images are free from aurora, 12.8\% with contamination at
high-airmass areas, 22.9\% with contamination at low-airmass areas,
and 9\% with aurorae all over the sky.  This is consistent with the
above quantitative Nigel results, although the data were taken in
different years.

It has been noticed from KLCAM data that the intensity of aurorae has
a very wide distribution, therefore weak aurorae probably occur more
frequently than what naked eyes can detect.  Quantitative analyses on
the data are necessary and expected, and monitoring should continue to
span multiple years.

\subsection{THz Transmission and PWV}
\label{sec:thz}
\label{sec:submm}

Site testing for terahertz or sub-mm (1\,THz $\equiv$ 0.3\,mm) is
less complicated because the ultimate determining factor for the
atmospheric transmission in this spectral region is the perceptible
water vapor.  Studies have confirmed Dome~A being the best
site on the ground for THz observations.

The first direct measurements of the THz atmospheric transmission
above Dome~A were made with Pre-HEAT at 661\,GHz (453\micron) in 2008.
The best 25\% atmospheric transmission in winter was 80\%,
corresponding to a PWV of 0.1\,mm, and daily averages could be as low
as 0.025\,mm.  This suggests that new far-infrared
spectral windows are opened at Dome~A for many sciences like star
formation, the life cycle of interstellar matter, and evolution of galaxies, etc.

By modeling the water absorption bands at 710--740\,nm and
800--840\,nm in the Nigel spectra and calibrating against
satellite data, \citet{Sims12b} managed to derive the best 25\% of PWV
of 0.09\,mm, consistent with Pre-HEAT results.  They also
demonstrated, through modeling, the benefits of the dry air to optical
and near-IR transmission between 700\,nm and 2.5\micron.

The measurements of atmospheric radiation from 20\micron\ to
350\micron\ was carried out by FTS (Sect.~\ref{sec:fts}) in 2010--2011
and transmittances were derived for the whole spectral region.
Substantial transmission in atmospheric windows throughout the whole
band was revealed, such as the windows above 7.1\,THz with
transmittance greater than 40\%.  Moreover, the time fraction of the
extremely dry conditions, thus opening the observing windows, is much
greater than at other sites where observations in these windows can be
attempted \citep{Shi16}.
Based on current modeling, the derived median PWV is 0.16\,mm in winter
(April to September) and 0.19\,mm for the entire year (Qijun Yao and
Zhenhui Lin, {\it private communications}).

Given that ``Ridge A'' is in the vicinity of Dome~A and its similar
altitude, the dry conditions there are expected to be similar or even
slightly better than at Dome~A based on estimates with satellite data
\citep{Saunders09}.  This is yet to be verified by the actual
measurements and results from HEAT (Sect.~\ref{sec:preheat})
\footnote{\it \url{http://soral.as.arizona.edu/HEAT/data/index.html}}.

\section{Instrumentation}
\label{sec:instrumentation}

Although the atmospheric properties favor Dome~A to be the best site
on earth for optical, IR, and THz astronomical observations,  there
remain great challenges in instrumentation as well as logistics, some
are even comparable to space.  Tremendous efforts have been made to
address the issues that include low temperature, easily icing/frosting
due to saturated water vapor in the air, limited power supply, limited
bandwith of communication, etc.  Most of the difficulties were known
from earlier experiments in Dome~C and South Pole and they were taken
into account for development of instruments.  \\

\noindent
{\it Conquer low temperature}

PLATO and PLATO-A provided a thermal-controlled space in the IM for
electronics as well as small instruments that could look through the
windows on the roof (e.g., Gattini, Nigel, Pre-HEAT, FTS, etc.).  For
others, they could be placed outside on the roof (e.g., HRCAM, KLCAM
etc.) or had to be installed on the snow surface separately (e.g.,
CSTAR, AST3, Snodar, KL-DIMM, etc.).

As the first experiment, CSTAR did not have any moving parts to avoid
possible mechanical failure, so it did not have a focusing mechanism.
To deal with focus change due to the huge temperature difference
between +30\degrc\ during assembly and $-$80\degrc\ at Dome~A, the main structure that supports
its optical system was made of INVAR36 which has very low coefficient
of thermal expansion \citep[CTE,][]{Yuan08}.  This was also applied to
AST3 \citep{Yuan14,Yuan16,Lixy19}.
For the same reason, KL-DIMM selected a telescope tube made of carbon
fiber to work between $-$30\degrc\ and $-$80\degrc\ as its focus could
be adjusted on-site in summer \citep{Ma18b}.

Low temperatures can easily cause mechanical failures, such as stuck
moving parts, because materials of different CTEs are used for gears
or the gear grease is not of the grade for temperature low enough.
This often happens with commercial products and can be fixed with
applying proper grease and upgrading some parts made with the same
material.  These approaches have been successful for the AP\,1600 mount of
KL-DIMM \citep{Ma18b}.

To make sure the moving parts can work at Dome~A, it is very necessary
to test in a cold chamber all moving components as well as the whole
instrument if size permits.  Cooling a part down to a temperature and
then checking its functionality at room temperature only verifies its
storage temperature, not working temperature, as was a mistake made
with the wind turbine experiment.
Therefore, simulating the actual operation of instruments at a winter
temperature of Dome~A is extremely important as it will reveal the
weak links.

Cold tests should also {be }applied to computers or other
electronic equipment down to $-20$\degrc\ if placed inside PLATO or
to the lowest temperature in the specification if installed outside.
For example, the AST3 CCD controller was tested to work well at
$-55$\degrc\ as specified, but it would be damaged at a lower
working temperature. Therefore, thermal protection must be provided,
usually using insulation materials to keep it at a working
temperature with heat generated from its own operation.  This has
been a common practice at Dome~A.  However, a danger in this
practice is that, if there was an extended power outage,  the
temperature of the electronic equipment could drop below its
specification and a cold start would damage it.

To solve the above problem, an active thermal control system with
temperature sensors was introduced and it has an external heating
component to warm up the equipment to its working temperature before
turning it on after it stopped work for a while.
The active thermal control system also has an internal heating
component that can be modulated for the equipment to work in a proper
temperature range, also avoiding overheating.
This active approach worked perfectly for KLCAM, the electronic boxes
of KLAWS and KL-DIMM, and CODS even during power outages in winter
\citep{Shang16,Shang18}.

In many cases, laptop computers had to be used
outside close to instruments at a temperate down below
{$-30$\degrc} and often shut themselves down for
protection in a very short period if not kept warm. It has been verified that it
is actually the battery that cannot withstand the low temperature,
so a modification of the power supply to laptops made the work
outside much easier and more efficiently during the 35th CHINARE
(Keliang Hu, {\it private communications}).

For future large telescopes, vertical temperature gradient also has to
be addressed in order to keep an ideal shape of a mirror or an antenna.
\\

\noindent
{\it Defrost and deice }

Although the PWV at Dome~A is very low (Sect.~\ref{sec:thz}), the
relative humidity is usually close to 100\% \citep{Hu18}.
Because of the saturated water vapor in air and frequent temperature
fluctuation, it is common for surfaces to {suffer}
from frosting in Antarctica.
Frosting on mirrors or glass windows exposed to open air was known,
but cannot be simulated in a cold chamber.  ITO conductive coating was
used, once turned on, to generate heat to sublimate frost or prevent it
from building up.
This worked greatly for CSTAR and KL-DIMM, but not very efficiently
for AST3, possibly because of the much larger area of its aperture.
Other approaches were developed for AST3, such as the inner and outer
air blowers, but the blowers could not be replaced if failed and
negative tube seeing could arise if turned on.  So more efficient
defrosting methods need to be explored.

Ice was found in between the teeth of AST3 gears and they are
considered the main reason of the RA or Dec axis getting stuck.
{The} ice could be the compressed frost or blowing
snow, but there was not enough power of heat to remove it in winter.
The current approach was to make good gear covers to prevent snow
being blown in as much as possible, but frosting could still occur
as it was not possible to make airtight covers because of wiring.
This worked well only in 2017, but improvements or new approaches are
still needed.

Careful thermal design can also help to solve the frosting problem.
KLCAM is a good example with intensive thermal control designs
\citep{Shang18}.  Besides a low-power active thermal control system
mentioned above, the inner structure of KLCAM was designed to guide
any heat generated inside only to the lens at the top and therefore
prevent any frost from building up.  This kind of designs certainly
works for small instruments, but the design concept can also be
learned for large ones.  \\

\noindent
{\it Cope with low power and low bandwidth}

PLATO-A  can provide 1\,kW in average in winter to support everything
on-site.  Not planning the power budget carefully would cause
struggled operations which had happened.
Any purchased devices and equipment must be selected to run on low
power if available, and customized equipment have to take every watt
into consideration, such as the data storage array in CODS which turns
on only one disk at a time to save power and also for data safety
\citep[][Sect.~\ref{sec:ast3}]{Shang12,Shang16}.

With Iridium Openport, an affordable plan can provide data of only
hundreds of MByte per month.  This is far from enough for transferring
image data from CSTAR, AST3, or even KL-DIMM.  Therefore, real-time
pipelines were developed and only useful results can be retrieved
promptly from Dome~A.  The most complicated pipeline so far is for
AST3 \citep{Ma20b}.

The monitoring and operational data for instruments can be transferred
back with scientific results.  However, since the bandwidth over
Iridium is limited, expensive, and sometimes unstable, a dedicated
narrow-band file transfer protocol was developed and implemented
\citep{Huangsy20}.
This demonstrates one of numerous examples of customized solutions for
astronomy from Dome~A.
\\

\noindent
{\it Unattended operation with redundancy}

It must be emphasized that instruments operated at Dome~A are not just
robotic, they are unattended for at least a year until the next
traverse team arrives, and are prone to single points of failure in
the harsh environment.  Therefore, reliability ultimately determines
the success of an instrument or device.
Among the techniques in reliability engineering, avoidance of single
point of failure is crucial in this situation and redundancy is an
effective solution.

Redundancy was taken into account since the beginning of CSTAR
project.  Four telescopes served as backups for each other and were
complementary with different filters if all had worked well.

For an individual instrument, such as one AST3 telescope, redundant
devices and equipment were provided wherever it was possible.  An
example of success was the CODS system which had an identical backup
for each of its three subsystems (Sect.~\ref{sec:ast3}).  Even for the
single CCD camera and controller, redundant operating system was
developed to enable image acquisition from two computers by splitting
the signals over optical fiber from one CCD to two acquisition cards
\citep{Shang16}.  This is certainly not necessary for a normal
observatory.  However, this technique had saved the 2017 observing
season when one acquisition card was damaged by cosmic rays early in
the season.

To achieve fully automatic and unattended operation of instruments,
customized software are inevitable, not just for pipeline, but also for
observing planning \citep{Liuq18}, data acquisition and storage,
coordination of devices \citep{Hu16} and communication with the
telescope \citep{Lixy13}.
A complicated software suite was developed for AST3 sky survey
\citep{Hu16} and it can be generalized for any robotic, unattended
observatory \citep{Hu18}.

To aid the operation and sky survey, monitoring information including
the status of instruments were always made online in real-time.
Remote, manual operation was possible and available, but it is the
last resort for a telescope at Dome~A.  A summary of automation of
AST3, the representative instrument, can be found in \citet{Ma20b}.
\\

\noindent
{\it Unstable ice sheet}

There remains a very important, but currently neglected problem that
will challenge future instrumentation at Dome~A.  \citet{Zhou13}
found that the south celestial pole moved around in CSTAR images
although CSTAR had no moving parts and was supposed to stare at the
south celestial pole all the time.  This indicates that the entire ice
sheet is moving around.

The 2008 CSTAR images have shown that the moving is roughly random
during a day and can be up to 45\arcsec; while for the whole observing
season in 2008, the locus covered a range up to 5\arcmin\ with a
preferred direction.

This did not affect either CSTAR or AST3 very much as they all have
very large FOVs.  However, for future large telescopes and/or
spectrographs that require accurate pointing, this issue has to be
studied carefully and taken into consideration.


\section{Future}

Over more than ten years, Dome~A has been gradually confirmed to be
the best site on earth for optical, infrared, and THz
astronomical observations.   To take advantage of the unique resources
sooner than later, large projects have been proposed.

Kunlun Dark Universe Survey Telescope (KDUST) will be a 2.5\,m
optical/IR telescope with a large FOV of 1.5\degr$\times$1.5\degr\ and
an image quality of 0.2\arcsec\,FWHM to match the free-atmosphere
seeing \citep[e.g.,][]{Zhu14}.  It is expected to be able to compete with 6\,m
class telescopes at the best mid-latitude sites.
The main research areas by KDUST survey include dark energy, dark
matter, Galactic structure, and exoplanets, etc.
\citep[e.g.,][]{Zhao11}.

The 5\,m Dome A TeraHertz Explorer (DATE5) will work at
wavelengths around
350\micron\ (FOV 10\arcmin$\times$10\arcmin) and
200\micron\ (FOV 5\arcmin$\times$5\arcmin) in the THz frequencies
\citep[e.g.,][]{Yang13}.
These wavelengths are chosen based on the dry conditions at Dome~A
and also for studies of
starburst galaxies and star formation,
interactions between black holes and stellar/interstellar matter,
life cycle of matter in the near universe,
molecular clouds in Milky Way,
and origin of solar system, etc.

These two telescope{s} are also planned to be unattended before
Kunlun Station becomes a winterover station.  The demands for
logistics including power are huge but manageable.  The construction
is expected to take 5 years after the projects are approved.

In the meantime, the current site testing instruments will still be
operating not just to collect more monitoring data, but also provide
long-term assessment of the site and critical information for the
design, construction, and operation of KDUST and DATE5.

Further site testing {is} still needed because
parameters like the vertical turbulence profile (\cntwo), and thus
the coherence time and isoplanatic angle, etc. are still not clear,
but these need to be understood for future adaptive optics and
interferometry applications.
Relevant instruments have been proposed such as the Multistar
Turbulence Monitor \citep[][Sect.~\ref{sec:seeing}]{Hickson19}.
Moreover, A Lunar Scintillometer \citep{Hickson04} called KunLun
Turbulence Profiler (KLTP) is being developed.  KLTP will have
multiple photo-diodes arranged linearly with different separations to
measure scintillation from the Moon.  The correlations between
intensities from the pairs of detectors with different baselines can
lead to the measurements of turbulence at different heights.

Daytime observations of the Sun and bright stars
will continue to explore more potential of Dome~A,
and AST3-2 could be revived.

The expected low IR sky background are yet to be confirmed with
measurements at Dome~A.  Nevertheless, a new project named Kunlun
Infrared Sky Survey (KISS) was proposed to equip an infrared camera to
the third AST3 telescope AST3-3.
KISS also plans to carry out time-domain studies, but specifically
in the 2.36\micron\ {$K_{\rm dark}$} band which
benefits very low background in Antarctica.
For this project and also for solving the problem of heat dissipation
inside the telescope tube, AST3-3 has a modified design to move the
focal plane outside its tube by adding a fold mirror.
With a 2k$\times$2k HgCdTe detector, the IR camera will have a plate
scale of {1.35\arcmin~pixel$^{-1}$} and
an FOV of 46\arcmin$\times$46\arcmin\ \citep{Lawrence16}, and it is
being built at Australian Astronomical Observatory based on the
collaboration between China and Australia.


Depending on the outcome and experience from KDUST and DATE5,
even larger projects are expected, but this has to develop with the
logistic support from Kunlun Station which will evolve into
a winterover station.




\begin{acknowledgements}
The author is grateful to the CHINARE teams supported by CAA and
PRIC.
The author thanks
M. Ashley,
K. Hu,
Y. Hu,
P. Jiang,
Y. Li,
Z. Lin,
B. Ma,
X. Pang,
J. Wang,
S. Wang,
X. Yang,
Q. Yao,
for materials and helpful comments.
We are also thankful for AAS
and Springer Nature
for granting us permission
for figure reproduction.
Z. Shang has been supported in part by the National Natural Science
Foundation of China (Grant Nos.\,11733007, 11673037 and 11273019) and
the National Basic Research Program (973 Program) of China (Grant No.\
2013CB834900), the Chinese Polar Environment Comprehensive
Investigation $\&$ Assessment Program (Grant No.\ CHINARE2016-02-03).
\end{acknowledgements}

%


\end{document}